\documentclass[12pt,hidelinks]{article}

\usepackage{newtxtext,newtxmath}

\usepackage{graphicx}

\usepackage[letterpaper,margin=1in]{geometry}

\usepackage[T1]{fontenc}

\renewenvironment{abstract}
	{\quotation}
	{\endquotation}

\date{}

\makeatletter
\renewcommand{\fnum@figure}{\textbf{Figure \thefigure}}
\renewcommand{\fnum@table}{\textbf{Table \thetable}}
\makeatother

\usepackage{scicite}

\usepackage{url}

\usepackage{doi}

\def\scititle{
    A quadratic-scaling  algorithm\\ with guaranteed convergence \\for quantum coupled-channel calculations
}
\title{\bfseries \boldmath \scititle}

\author{
	Hubert J. J\'{o}\'{z}wiak$^{1,2}$,
	Md Muktadir Rahman$^{3}$,
	Timur V. Tscherbul$^{3}$$^\ast$\and
	\small$^{1}$Institute for Molecules and Materials, Radboud University, Nijmegen, The Netherlands \and
	\small$^{2}$Institute of Physics, Faculty of Physics, Astronomy and Informatics,\and
\small Nicolaus Copernicus University in Toru\'{n}, Grudzi\k{a}dzka 5, 87-100 Toru\'{n}, Poland.\and
	\small$^{3}$Department of Physics, University of Nevada, Reno, NV, 89557, USA.\and
	\small$^\ast$Corresponding author. Email: ttscherbul@unr.edu; timur.v.tscherbul@gmail.com \and
}

\begin{document} 

\maketitle

\begin{abstract} \bfseries \boldmath
Rigorous quantum dynamics calculations provide essential insights into complex scattering phenomena  across atomic and molecular physics, chemical reaction dynamics, and astrochemistry. However, the application of the gold-standard quantum coupled-channel (CC) method has been fundamentally constrained  by a steep cubic scaling of computational cost $[\mathcal{O}(N^3)]$. Here, we develop a general, rigorous, and robust method for solving the time-independent Schrödinger equation for a single column of the scattering S-matrix with
 quadratic scaling $[\mathcal{O}(N^2)]$ in the number of channels. The Weinberg-regularized Iterative Series Expansion (WISE) 
algorithm resolves the divergence issues affecting iterative  techniques 
 by applying a regularization procedure to the kernel of the multichannel Lippmann-Schwinger integral equation.
 The method also explicitly incorporates closed-channel effects, including those responsible for multichannel Feshbach resonances. We demonstrate the power of this approach by performing rigorous calculations on He~+~CO and CO~+~N$_2$ collisions, achieving exact quantum results with 
 quadratic scaling
{guaranteed by a contour-integral  construction.} Our results establish a highly scalable computational paradigm, enabling state-to-state quantum scattering computations for complex molecular systems.
 
\end{abstract}

\section*{Introduction}
\noindent
Quantum collision dynamics of atoms, molecules, and nuclei is central to a vast array of fields ranging from atomic, molecular, and optical (AMO) physics \cite{Chin:10,Greene:17,Book_Krems} to chemical reaction dynamics \cite{Althorpe:03,Clary:08,Krems:08,Balakrishnan:16}, astrochemistry \cite{Roueff_2013,Dishoeck:13}, and nuclear physics \cite{Hagino:22}. Precision calculations of scattering observables -- such as state-to-state cross sections and reaction rates -- provide essential {quantitative} insights into complex quantum phenomena abundant in these fields. Examples include intricate mechanisms of chemical reactions \cite{Althorpe:03}, the broadening of spectral lines in the atmospheres of Earth \cite{Hartmann2008,Rayer_2020,Jozwiak_2021,Olejnik_2023}, Solar System planets \cite{Jozwiak2024} and exoplanets \cite{Tennyson_2017, Niraula_2022, Chubb_2024,Wiesenfeld_2025},  the fate of water and small polyatomic molecules in the interstellar medium \cite{Dishoeck:13,Roueff_2013}, and the mechanisms of ultracold molecular collisions  \cite{Krems:08,Balakrishnan:16,Bohn:17,Vogels:15,Bause:23,Margulis:23,Horn:25,Tang:23,Park:23b,Morita:24,Liu:25,Soley:25} and atomic few-body recombination  \cite{Greene:17,DIncao:18,Haze:25,Higgins:25}. A key challenge across these fields is the need to rigorously model ever-larger, more complex systems that remain out of reach of current computational methods.

Time-independent coupled-channel (CC) calculations are the established gold standard for elucidating quantum collision dynamics in AMO physics, chemical reaction dynamics, and astrochemistry \cite{Althorpe:03,Clary:08}. CC methods are uniquely suited for precision calculations of state-to-state observables \cite{Skouteris:00}, particularly at the low collision energies relevant for ultracold chemistry and astrochemical modeling. However, the rigorous application of CC methodology is fundamentally limited by a steep cubic scaling [$O(N^3)$] of computational cost with the number of collision channels, $N$. 
While powerful time-dependent wavepacket methods do achieve quadratic scaling [$O(N^2)$], they are inherently ill-suited for calculating state-to-state collision dynamics at low collision energies due to the extensive spatial grids and propagation times required.
As a result, complex quantum phenomena of major fundamental and applied importance, such as H$_2$O~+~H$_2$O or H$_2$O~+~CH$_3$OH collisions \cite{Bostan:24}, remain completely intractable using rigorous quantum scattering methodology.

Early efforts to bypass the cubic scaling barrier, most notably by Thomas \cite{Thomas_1979,Thomas_1982}, demonstrated that the CC equations could be formally solved for a single column of the scattering $S$-matrix with $\mathcal{O}(N^2)$ complexity. This reduction from cubic to quadratic scaling is possible in principle 
because the dynamical information contained within a single $S$-matrix column is sufficient to yield complete state-to-state scattering observables for molecules prepared in a specific quantum state—a scenario ubiquitous in ultracold chemistry, astrochemistry, and crossed-molecular beam collision experiments. However, despite its promise, Thomas’s iterative approach \cite{Thomas_1979,Thomas_1982} suffers from two critical limitations that have likely precluded its widespread adoption. First, it lacks a mechanism to incorporate closed channels, which are essential for numerical convergence and for properly capturing threshold and Feshbach resonance phenomena. Second, and more fundamentally, the underlying iterative scheme frequently diverges when applied to the deep and highly anisotropic interaction potentials characteristic of complex molecular systems \cite{Thomas_1979,Thomas_1982}.


Here, we overcome these long-standing methodological barriers to deliver the first robust    
quadratic-scaling algorithm 
for solving the time-independent Schr\"odinger equation for a single column of the scattering $S$-matrix. Our approach is based on the iterative solution of the multichannel Lippmann-Schwinger (LS) equation and introduces two fundamental innovations. First, unlike previous iterative techniques, it rigorously incorporates closed channels, which are essential for numerical convergence of scattering observables.
Second, to eliminate long-standing divergence issues affecting previously developed iterative expansions \cite{Thomas_1979,Thomas_1982,RhoadesBrown:80}, we formulate a general regularization procedure  in which the Weinberg eigenvalues
of the LS kernel that cause the iterative series to diverge are identified and removed. 

The resulting {\it Weinberg-regularized Iterative Series Expansion (WISE)} algorithm guarantees numerical convergence of scattering observables regardless of the strength or anisotropy of the interaction potential.
We demonstrate the robustness and quadratic scaling of the WISE algorithm by applying it to multichannel He~+~CO and CO~+~N$_2$ collisions on {\it ab initio} potential energy surfaces. These results enable rigorous quantum scattering calculations on a wide array of molecular systems previously considered intractable, contributing to our fundamental understanding of the intricate quantum dynamics of complex molecular collisions and paving the way for large-scale numerical simulations in ultracold molecular physics, chemical reaction dynamics and astrochemistry.

\vspace{-0.5cm}
\section*{Results}
\vspace{-0.3cm}
\subsection*{Theory}
Our starting point is the time-independent Schrödinger equation for the scattering of two molecules, $\hat{{H}} \Psi = E \Psi$. The Hamiltonian of the collision complex is given by
\begin{equation}
\label{eq:H}
    \hat{H} = -\frac{\hbar^{2}}{2\mu} \nabla_{\vec{r}}^{2} + \hat{H}_{\mathrm{int}}(\boldsymbol{\xi}) + \hat{V}(\vec{r},\boldsymbol{\xi}),
\end{equation}
where $\mu$ is the reduced mass of the complex, $\vec{r}$ is the vector connecting the centers of mass of the colliding molecules, $\boldsymbol{\xi}$ denotes all internal coordinates, and $\hat{V}$ is the interaction potential. The term $\hat{H}_{\mathrm{int}}$ describes the internal structure of the isolated monomers.

In the framework of the coupled-channel (CC) approach, the total wavefunction $\Psi$ is expanded in a complete basis of scattering channels \cite{Arthurs_1960}
\begin{equation}
\label{eq:Psi}
    \Psi^{JM_{J}}( \vec{r},\boldsymbol{\xi}) = \frac{1}{r} \sum_{\gamma, L} u_{\gamma L}^{JM_{J}}(r) \mathcal{Y}_{\gamma L}^{JM_{J}} (\hat{r},\boldsymbol{\xi} ),
\end{equation}
where $r$ is the magnitude of $\vec{r}$, and $\hat{r} = \vec{r}/r$ specifies its orientation in space. The channel basis functions $ \mathcal{Y}_{\gamma L}^{JM_{J}} (\boldsymbol{\xi}, \hat{r})$ are 
simultaneous eigenfunctions of the internal Hamiltonian, $\hat{H}_{\mathrm{int}}$ (indexed by $\gamma$), the end-over-end rotational angular momentum $\hat{{L}}^{2}$ (indexed by $L$), and the total angular momentum
squared $\hat{J}^2$. The total angular momentum of the collision pair ${J}$ and its projection on the quantization axis $M_{J}$ are good quantum numbers in the absence of external fields. For the specific case of atom-molecule collisions, the internal index $\gamma$  corresponds to the rotational angular momentum of the molecule, denoted as $j$. For collisions between two molecules, $\gamma \equiv \{j_{\mathrm{A}},j_{\mathrm{B}},j_{\mathrm{AB}}\}$  
comprises individual rotational angular momenta ($j_{\mathrm{A}}, j_{\mathrm{B}}$) and their vector sum  ($j_{\mathrm{AB}}$).

Substituting the expansion \eqref{eq:Psi} into the Schrödinger equation 
yields the standard set of CC equations for the radial expansion coefficients, $u_{\gamma L}^{JM_{J}}(r)$ 
\begin{align}
\begin{split}
    \label{eq:CC_diff} 
    \Bigl(  \frac{d^2}{dr^2} + k_{\lambda}^2 - \frac{L_{\lambda}(L_{\lambda}+1)}{r^2} - U_{\lambda \lambda}(r) \Bigr) u_{\lambda}(r) = \sum_{\lambda' \neq \lambda} U_{\lambda \lambda'}(r) u_{\lambda'}(r). 
\end{split}
\end{align}
Here, we have introduced a collective channel index $\lambda \equiv \{\gamma, L\}$ and suppressed the $J$ and $M_{J}$ labels for brevity. Further, $k_{\lambda}^2 = 2\mu(E - E_{\lambda})/\hbar^2$ is the squared wavevector for channel $\lambda$ and $\mathbf{U}(r) = (2\mu/\hbar^2)\mathbf{V}(r)$ is the scaled interaction potential matrix.

In Eq.~\eqref{eq:CC_diff}, the terms on the left-hand side define the reference Hamiltonian, $\hat{H}_{\lambda}$, for each channel, which includes the centrifugal term and the diagonal part of the interaction potential. The sum on the right-hand side contains couplings ($U_{\lambda \lambda'}$) between different channels. In practice, the infinite expansion in Eq.~\eqref{eq:Psi}, and the summation in Eq.~\eqref{eq:CC_diff} are truncated to include only a finite set of $N$ strongly coupled channels to ensure convergence.

Standard methods solve the CC equations \eqref{eq:CC_diff}  by propagating a matrix of $N$ linearly independent solution vectors outwards from the classically forbidden region \cite{Johnson_1973, Johnson_1978, Manolopoulos_1986}. Matching this matrix to the proper boundary conditions yields the full scattering matrix $\mathbf{S}$. Because such propagation involves matrix-matrix operations at every grid step, the computational cost scales as $\mathcal{O}(N^{3})$, effectively limiting solvable problems to a few tens of thousands of channels---to our knowledge, the largest published CC calculation to date involved $N\simeq 18$,850 channels \cite{Suleimanov:12}.


Instead of solving the system of coupled second-order differential equations \eqref{eq:CC_diff} directly, as done in previous work \cite{Johnson_1973,Johnson_1978,Manolopoulos_1986}, we recast  it in the form of the Lippmann-Schwinger (LS) integral equation \cite{Newton_2013,Thomas_1979,Thomas_1982} for a single-column solution vector $\vec{u}(r)$ corresponding to a specific incoming channel $\lambda_{0}$
\begin{equation}
    \label{eq:LS}
    \vec{u}(r) = \vec{u}_{0}(r) + \mathbf{K}\vec{u}(r),
\end{equation}
which can be solved iteratively using the standard Born series, $\vec{u}(r) = \sum_{n=0}^{\infty} \mathbf{K}^n \vec{u}_{0}(r)$ \cite{Newton_2013}. 
Importantly, because the iterative computation involves only matrix-vector operations ({\it i.e.}, the repeated application of $\mathbf{K}$ to $\vec{u}_{0}$), it scales quadratically as $\mathcal{O}(M N^{2})$, where $M$ is the number of iterations. For $M \ll N$, this offers a substantial advantage over standard direct propagation techniques \cite{Johnson_1973,Johnson_1978,Manolopoulos_1986}.  However, as noted above, the naive iterative approach---originally developed by Thomas \cite{Thomas_1979, Thomas_1982}---suffers from two major limitations that have likely hindered its widespread use: (i) the inability to rigorously account for closed-channel effects, and (ii) the divergence of the Born series for strong or attractive potentials. 
Here, we describe how the WISE framework overcomes these challenges through the robust, divergence-free inclusion of closed channels and the spectral regularization of the  kernel matrix $\mathbf{K}$.


In Eq.~\eqref{eq:LS}, the unperturbed source term $\vec{u}_{0}(r)$ imposes the incoming-wave boundary condition; its components are $u_{0, \lambda}(r)=\delta_{\lambda {\lambda_{0}}}x_{\lambda}(r)$, where $x_{{\lambda}}(r)$ is the regular solution of the single-channel Schr\"odinger equation for the reference Hamiltonian $\hat{H}_{\lambda}$
\begin{equation}
\label{eq:single}
    \Bigl(  \frac{d^2}{dr^2} + k_{\lambda}^2 - \frac{L_{\lambda}(L_{\lambda}+1)}{r^2} - U_{\lambda \lambda}(r) \Bigr) x_{{\lambda}}(r) = 0.
\end{equation}
with the boundary condition $\lim_{r\to 0}x_{\lambda}(r) \to 0$. Its asymptotic form depends on the channel energy. For open channels ($k_{\lambda}^{2} > 0$), it takes the form:
\begin{align}
    \begin{split}
    \label{eq:x_asympt}
        x_{\lambda}(r) &  = k_{\lambda}r \Bigl(\cos{\phi_{\lambda}} {j}_{L}(k_{\lambda}r) - \sin{\phi_{\lambda}} n_{L}(k_{\lambda}r)\Bigr) ,
    \end{split}
\end{align}
which is appropriate when the interaction potential is negligible compared to the centrifugal term, and involves the spherical Bessel ($j_{L}$) and von Neumann ($n_{L}$) functions; $\phi_{\lambda}$ is the elastic phase shift for the diagonal potential $U_{\lambda \lambda}(r)$. Since the Schrödinger equation is a second-order differential equation, there exists a second, linearly independent solution to Eq.~\eqref{eq:single}, $y_{\lambda}(r)$, diverging at the origin and behaving asymptotically as
\begin{align}
    \begin{split}
        \label{eq:y_asympt}
        y_{\lambda}(r) & =  -{ie^{i\delta_{\lambda}}}  r \Bigl(j_{L}(k_{\lambda}r) + i n_{L}(k_{\lambda}r)\Big).
    \end{split}
\end{align}
The action of the integral operator $\mathbf{K}$ in Eq.~\eqref{eq:LS} is defined by the kernel
\begin{align}
    \label{eq:kernel}
    \mathbf{K} \vec{u}(r)  =\int_{0}^{\infty} \mathbf{G}(r,r')\mathbf{U}^{\mathrm{off}}(r')\vec{u}(r')\mathrm{d}r' .
\end{align}
Here, the matrix $\mathbf{U}^{\mathrm{off}}$ contains only the off-diagonal elements of the potential matrix, and $\mathbf{G}(r,r')$ is the Green's function matrix constructed explicitly from the reference solutions of Eq.~\eqref{eq:single}:
\begin{equation}
\label{eq:G_def}
    G_{\lambda \lambda'}(r,r') = \delta_{\lambda \lambda'} x_{\lambda}(r_{<}) y_{\lambda}(r_{>}),
\end{equation}
where $r_{<} = \mathrm{min}(r,r')$ and $r_{>} = \mathrm{max}(r,r')$. In practice, the radial coordinate is discretized onto a grid of $N_{\mathrm{g}}$ points, $\{r_{1}, r_{2},\dots,r_{N_{\mathrm{g}}}\}$. This discretization transforms the integral operator $\mathbf{K}$ into a non-symmetric matrix of dimension $(N\cdot N_{\mathrm{g}}) \times (N\cdot N_{\mathrm{g}})$:
\begin{equation}
    \label{eq:K_matrix}
    \mathrm{K}_{\alpha\beta } = \mathrm{K}_{(\lambda,i)(\lambda',i') }  = G_{\lambda\lambda}(r_{i},r_{i'}) U^{\mathrm{off}}_{\lambda\lambda'}(r_{i'}) w_{i'},
\end{equation}
where $\alpha$ and $\beta$ are composite indices flattening the channel and grid dimensions, e.g., $\alpha = (\lambda, i )\in \{1, \dots, N\cdot N_{\mathrm{g}}\}$, and $w_{i'}$ are the quadrature weights associated with the grid points. The sought-after vector $\vec{u}(r)$ determines the \textit{single column} of the $S$-matrix corresponding to the incoming channel $\lambda_{0}$ \cite{Thomas_1979,Thomas_1982}
\begin{align}
    \begin{split}
        \label{eq:Smatrix}
        S_{\lambda \lambda_{0}} = e^{i(\phi_{\lambda}+\phi_{\lambda_{0}})}\Bigl(\delta_{\lambda \lambda_{0}}  - \frac{2i}{\sqrt{k_{\lambda}k_{\lambda_{0}}}} \int_{0}^{\infty} x_{\lambda}(r) \sum_{\lambda' \neq \lambda}{U}^{\mathrm{off}}_{\lambda \lambda'}(r)u_{\lambda'}(r) \mathrm{d}r\Bigr) .
    \end{split}
\end{align}

In the following, we describe how the WISE framework overcomes the two primary challenges of the naive iterative approach \cite{Thomas_1979, Thomas_1982}: the rigorous inclusion of closed channels and the guaranteed convergence of the iterative series through spectral regularization.

\subsubsection*{Robust Integration of Closed Channels}


The first major limitation of the naive iterative approach is the omission of energetically inaccessible (closed) channels with $k_{\lambda}^{2}<0$.
In molecular scattering, these channels must be rigorously included in the basis set to achieve converged results and capture the full dimensionality of the interaction potential \cite{Child:74,Pack:87,Manolopoulos:90}.
Closed channels are also  responsible for fundamental scattering phenomena, such as Feshbach resonances \cite{Chin:10,Morita:24,Tang:23,Park:23b,Margulis:23,Horn:25}. In practice, including closed channels in CC calculations can be challenging because
their irregular wavefunctions 
 grow exponentially in the classically forbidden region,
  making standard propagation schemes notoriously unstable, and necessitating advanced techniques—such as those based on the log-derivative of the multichannel wavefunction \cite{Johnson_1973, Johnson_1978, Manolopoulos_1986}—to maintain numerical stability.
 


To overcome this limitation, we developed a stable ratio-propagation method for constructing the Green's function matrix $\mathbf{G}$. First, we solve the single-channel reference equation [Eq.~\eqref{eq:single}] by propagating the ratio of the \textit{regular} solutions $Q^{(x)}_{\lambda}(r_{i}) = x_{\lambda}(r_{i-1}) / x_{\lambda}(r_{i})$ outwards from the origin using the renormalized Numerov algorithm \cite{Johnson_1978}, starting with $Q^{(x)}_{\lambda}(r_{1}) =0$. For asymptotically open channels, we match this ratio to the asymptotic form in Eq.~\eqref{eq:x_asympt} to determine the elastic phase shift, $\phi_{\lambda}$.
Simultaneously, we determine the ratio of  \textit{irregular} solutions, $Q^{(y)}_{\lambda}(r_{i}) = y_{\lambda}(r_{i+1}) / y_{\lambda}(r_{i})$
by starting at the final grid point, $r_{N_\text{g}}$ using the asymptotic form of $y_{\lambda}$. For open channels, this is straightforward using Eq.~\eqref{eq:y_asympt}. For closed channels, however, the boundary conditions are completely different--- 
the regular and irregular solutions must behave as modified Bessel functions of the first ($I_{L+\frac{1}{2}}$) and second ($K_{L+\frac{1}{2}}$) kind:
\begin{align}
    \begin{split}
        x_{\lambda}(r) &= i^{L+1} \sqrt{\frac{\pi}{2}\kappa_{\lambda}r} \Bigl(\alpha_{\lambda} I_{L+\frac{1}{2}}(\kappa_{\lambda}r) + \beta_{\lambda} K_{L+\frac{1}{2}}(\kappa_{\lambda}r)\Bigr),\\
        y_{\lambda}(r)&= -\frac{i^{-(L+1)}}{\kappa_{\lambda}} \sqrt{\frac{2}{\pi}\kappa_{\lambda} r } \alpha_{\lambda}^{-1}K_{L+\frac{1}{2}}(\kappa_{\lambda}r),
    \end{split}
\end{align}
where $\kappa_{\lambda} = \sqrt{-k^{2}_{\lambda}}$. The regular solution must be finite at the origin, which fixes the ratio of the coefficients $\alpha_{\lambda}$ and $\beta_{\lambda}$. Direct evaluation of these coefficients in terms of $Q^{(x)}_{\lambda}$ is unstable due to the exponential growth of $I_{L+1/2}$.

To circumvent this problem, we derive a stable expression for the Green's function at the final grid point, $G_{\lambda \lambda}(r_{N_\mathrm{g}},r_{N_\mathrm{g}})$, that depends only on the ratio of the regular solution, $Q^{(x)}_{\lambda}$, and ratios and products of modified Bessel functions
\begin{equation}
    {G}_{\lambda\lambda'}(r_{N_\mathrm{g}},r_{N_\mathrm{g}}) = - \delta_{\lambda\lambda'}r_{N_\mathrm{g}} \left(   1 - \frac{Q^{(x)}_{\lambda}(r_{N_\mathrm{g}}) -\sqrt{\frac{r_{N_\mathrm{g}-1}}{r_{N_\mathrm{g}}}}\frac{I_{L+\frac{1}{2}}(\kappa r_{N_\mathrm{g}-1})}{I_{L+\frac{1}{2}}(\kappa r_{N_\mathrm{g}}) }}{Q^{(x)}_{\lambda}(r_{N_\mathrm{g}}) - \sqrt{\frac{r_{N_\mathrm{g}-1}}{r_{N_\mathrm{g}}}}\frac{K_{L+\frac{1}{2}}(\kappa r_{N_\mathrm{g}-1})}{K_{L+\frac{1}{2}}(\kappa r_{N_\mathrm{g}}) }} \right)
    I_{L+\frac{1}{2}}(\kappa r_{N_\mathrm{g}}) K_{L+\frac{1}{2}}(\kappa r_{N_\mathrm{g}}).
\end{equation}
Note that ${G}_{\lambda\lambda'}(r,r) \to -1/\kappa_\lambda $ as $r \to \infty$.
With the asymptotic value of $G_{\lambda\lambda}(r_{N},r_{N})$ established, the full Green's function matrix elements on the spatial diagonal ($r=r'$) are generated by inward propagation
\begin{align}
    \begin{split}
\label{eq:green_diagonal}
{G}_{\lambda\lambda'}(r_{i-1},r_{i-1})
 = \delta_{\lambda \lambda'}{Q}_{\lambda}^{(x)}(r_{i}) {G}_{\lambda\lambda'}(r_{i},r_{i}) ({Q}_{\lambda}^{(y)}(r_{i-1}))^{-1} .       
    \end{split}
\end{align}
Once the diagonal elements of $\mathbf{G}$ are known, the off-diagonal elements $(r\neq r')$ 
are evaluated using the separability of the Green's function:
\begin{align} \begin{split}
    \label{eq:green_offdiag}
    G_{\lambda\lambda}(r_{i},r_{i'}) 
    &= \begin{cases} G_{\lambda\lambda}(r_{i},r_{i}) \prod_{k=i'+1}^{i} Q^{(x)}_{\lambda}(r_k) & \text{for } i' < i, \\ G_{\lambda\lambda}(r_{i},r_{i}) \prod_{k=i+1}^{i'} Q^{(y)}_{\lambda}(r_k)^{-1} & \text{for } i' > i. \end{cases}
    \end{split}\end{align}
This procedure allows us to construct the full Green's function matrix without explicitly evaluating the exponentially growing wavefunctions, providing a robust approach for incorporating asymptotically closed channels into  iterative solvers of the LS equation.

Armed with this approach, we can now elucidate the 
effect of closed channels on the convergence of the iterative Born series. To this end,  consider a reduced-dimensional model of rotational energy transfer in He~+~CO collisions involving two open channels ($j=0,L=0$ and $j=1, L=1$), see Methods for details. The model describes rotational relaxation of CO molecules in collisions with $^4$He atoms at a collision energy of 5 cm$^{-1}$ and $J=0$.

The left panel of Fig.~\ref{fig:method}(b) shows the convergence of the $S$-matrix elements for  
the $j=1 \to 0$ collisional transition  computed using the iterative approach. We observe that in the absence of closed channels, convergence to the exact CC result is reached after $\simeq$10 iterations.
Strikingly,  adding a single closed channel ($j=2, L=2$) 
causes the Born series to diverge catastrophically. This instability is the second major limitation of iterative methods in quantum scattering theory \cite{Newton_2013}. We note that this divergence commonly occurs even in the absence of closed channels \cite{Thomas_1979,Thomas_1982} and can be mitigated by using, e.g., 
optimized initial guess vectors \cite{Thomas_1979}, Pad\'e approximants \cite{RhoadesBrown:80}, and block-diagonal Green's functions \cite{Thomas_1982}.
However, to our knowledge, none of these approaches guarantees convergence to the exact solution, particularly in the strong-interaction regime.

 


 \subsubsection*{Spectral Regularization and Guaranteed Convergence}

To address the second major limitation and ensure  convergence of the iterative Born series, we consider the spectral radius of the LS kernel, $\rho(\mathbf{K})=\mathrm{max}_{k}|\eta_{k}|$, where $\{\eta_{k}\}$ are the eigenvalues of $\mathbf{K}$~\cite{Newton_2013}. These eigenvalues, formally introduced in the context of nuclear scattering by Weinberg~\cite{Weinberg_1963}, and  recently revisited as a diagnostic tool for nucleon-nucleon interactions \cite{Hoppe:17}, serve as a valuable   indicator of the ``perturbativeness'' of the multichannel interaction potential. Specifically, the Born series converges only if $\rho(\mathbf{K})<1$; the presence of Weinberg eigenvalues  lying outside  the unit circle in the complex plane ($|\eta_{k}|>1$)  signals the divergence of the series.

We note that the number of these divergent Weinberg eigenvalues is always finite \cite{Weinberg_1963,Scadron:64}
provided the interaction potential operator $\hat{V}$ is short-ranged (true for most molecular and nuclear systems, excluding those described by Coulomb interactions). For these systems, the symmetrized scattering kernel $\hat{K}_S=\hat{V}^{1/2}\hat{G}\hat{V}^{1/2}$, which has the same eigenvalues as our LS kernel $\hat{K}$, is a compact operator  \cite{Scadron:64}. According to the Riesz-Schauder theorem \cite{ReedSimonBook},
the spectrum of a compact operator consists of a countable set of eigenvalues with no limit point in the complex plane except at zero. 
Thus, the number of divergent Weinberg eigenvalues approaches a constant  in the large $N$ limit{for a fixed scattering Hamiltonian. Fixing the Hamiltonian and converging with respect to the basis size $N$ is the standard approach in molecular quantum dynamics calculations \cite{Manolopoulos_1986}, which we adopt throughout this work. }

Figure~\ref{fig:weinberg} visualizes the structure of the discretized $\mathbf{K}$ matrix 
and its spectrum for the two and three-channel models of He~+~CO scattering. The matrix exhibits a distinctive asymmetric structure defined by the interplay between the Green's function and the interaction potential in Eq.~\eqref{eq:K_matrix}. Three distinct regions are visible in Figs.~\ref{fig:weinberg}(a) and \ref{fig:weinberg}(c). At very short range (column indices below 100), the matrix elements are small because the regular solution component $x(r)$ of the Green's function vanishes as $r\to 0$. At intermediate range (column indices  $\simeq 100{-}200$), a bright vertical strip appears where the interaction potential is strongest, and the Green's function is non-zero. Finally, at long range (column indices above 200), the matrix elements decay to zero as the potential decays, despite the oscillatory nature of the open-channel Green's function. 
The visible asymmetry---where columns, and not rows, dominate in the interaction region---arises because the interaction potential $\mathbf{U}^\text{off}(r')$ weighs the integration  variable ($r'$, columns) but not the observation variable ($r$, rows).

\begin{figure}[!ht]
    \centering
    \includegraphics[width=0.99\linewidth]{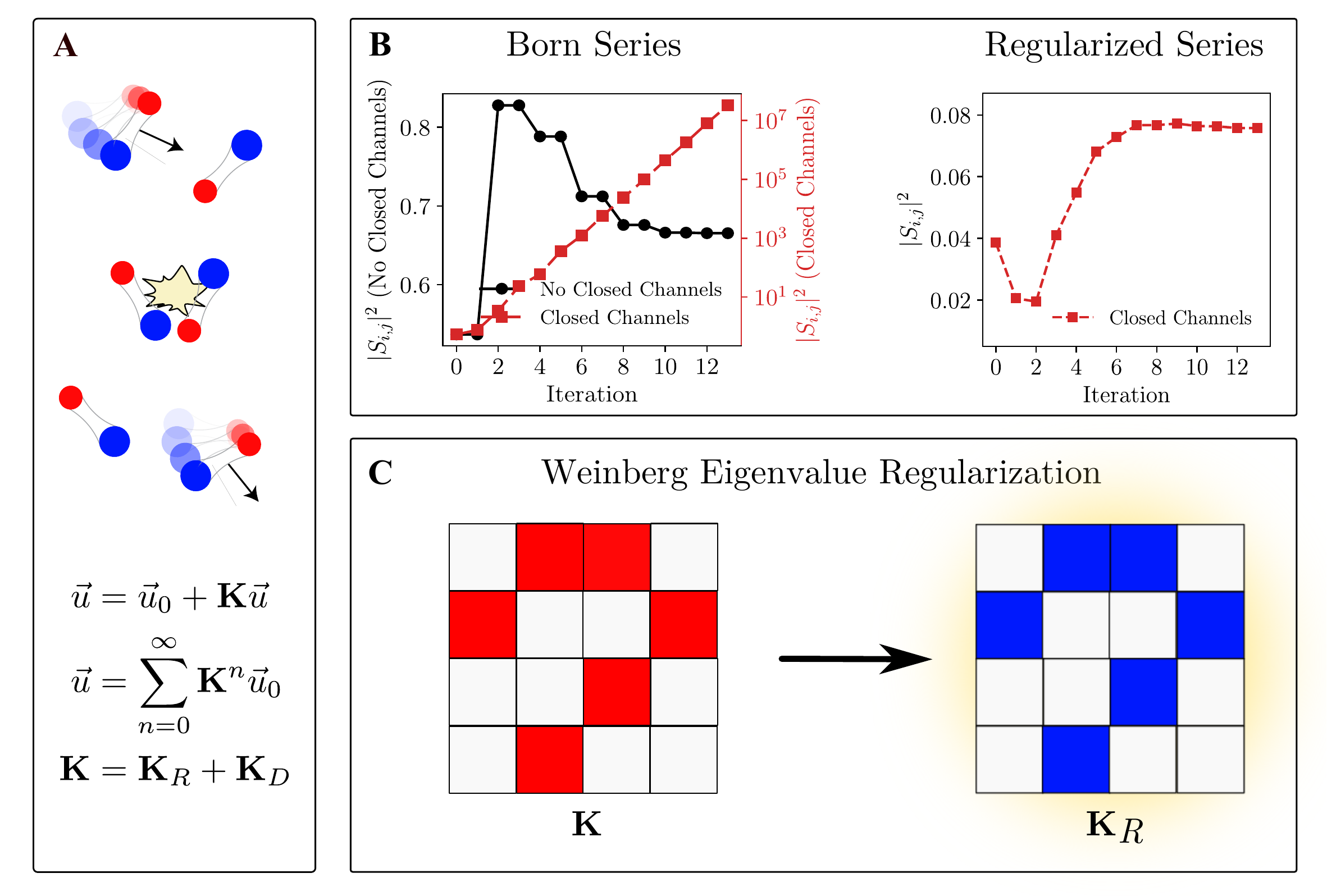}
    \caption{\textbf{Schematic of the WISE framework: Overcoming closed-channel divergence with spectral regularization.} (\textbf{A}) A pictorial representation of molecular collisions (top) and the key equations of the WISE method: the Lippmann-Schwinger equation for a single solution vector, $\vec{u}$, the standard Born series expansion, and the regularization procedure, where the kernel $\mathbf{K}$ is decomposed into a regular part $\mathbf{K}_R$ and a divergent part $\mathbf{K}_D$. \textbf{(B)}~Convergence of the squared S-matrix element $|S_{j=1,L=1; j=0,L=0}|^2$ for the reduced-dimensional  model of He~+~CO scattering. The left panel ("Born Series") compares the convergence of the Born series with only open channels (black circles, left axis, rapid convergence) against one with an added closed channel (red squares, right axis, logarithmic scale), where the series diverges catastrophically. The right panel ("Regularized Series") demonstrates that spectral regularization  restores stable convergence in the presence of closed channels. \textbf{(C)}~Conceptual schematic of the regularization procedure and key equations. The integral operator $\mathbf{K}$ is analyzed spectrally; "red" components represent the subspace of divergent eigenvalues (Weinberg eigenvalues with $|\eta| \ge 1$). These are projected out to form the regularized remainder $\mathbf{K}_R$ (blue blocks), which possesses a spectral radius within the unit circle, ensuring the convergence of the iterative series.}
    \label{fig:method}
\end{figure}

\begin{figure}
    \centering
    
    \includegraphics[width=0.99\linewidth]{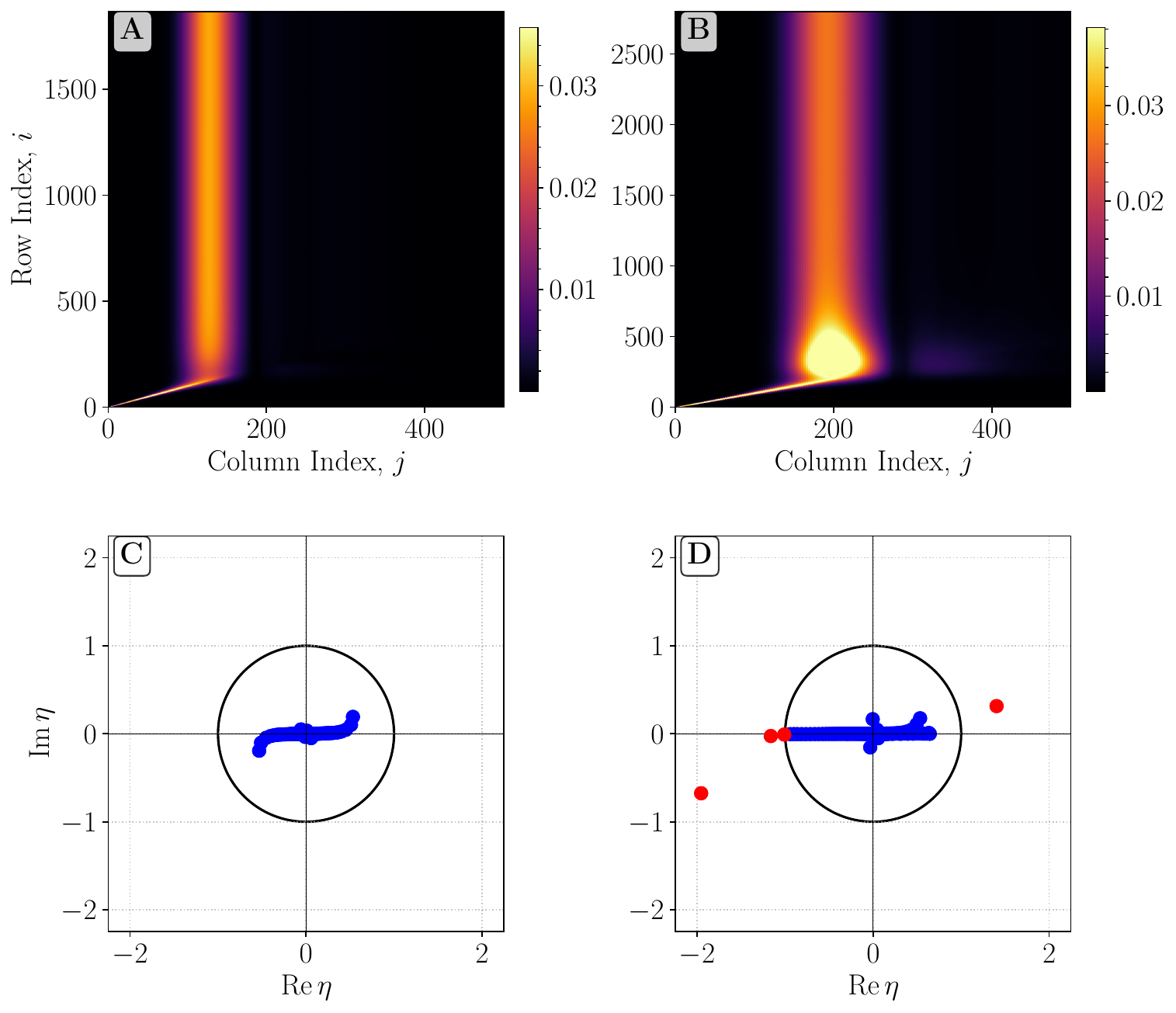}
    \caption{\textbf{Weinberg eigenvalue analysis of the kernel matrix $\mathbf{K}$. (A, B)} Structure of the discretized kernel matrix $\mathbf{K}$ for the reduced-dimensional He~+~CO model, plotted as absolute values $|\mathrm{K}_{\alpha \beta}|$. The axes represent composite indices flattening both channel and grid dimensions. The column axis is truncated at index 500 to highlight the relevant interaction region. The vertical stripe corresponds to the short-range region where the coupling potential is 
    {substantial}.
    \textbf{(C, D)} Corresponding spectra of Weinberg eigenvalues, $\eta_{k}$, in the complex plane. The solid black line indicates the unit circle. \textbf{(A,~C)} Two-open-channel model. All eigenvalues (blue dots) lie strictly within the unit circle ($|\eta_{k}| < 1$), ensuring convergence of the Born series. \textbf{(B,~D)} Three-channel model (two open channels plus one closed channel). Although the spatial structure of $\mathbf{K}$ appears qualitatively similar to (a), the inclusion of the closed channel pushes four eigenvalues (red dots) outside the unit circle ($|\eta_{k}| \geq 1$), causing   the standard iterative methods to diverge.}
    \label{fig:weinberg}
\end{figure}

In the two-channel case shown in Fig.~\ref{fig:weinberg}(c), this structure yields a spectrum entirely contained within the unit circle ($|\eta_k| < 1$), ensuring convergence of the Born series. In the three-channel case [see Fig.~\ref{fig:weinberg}(d)], the inclusion of the closed channel fundamentally alters the spectrum by pushing four Weinberg eigenvalues outside the unit circle ($|\eta_{k}| \geq 1$), and leading to the divergence observed in the right panel of Fig.~\ref{fig:method}(b). 
We attribute this to a {substantial}
enhancement   of the magnitude of $\mathbf{K}$-matrix elements at short range [Fig.~\ref{fig:weinberg}(b)] caused by the inclusion of the closed channel. 
The lack of convergence of the Born series is thus caused by the change in spectral properties of the kernel matrix $\mathbf{K}$ introduced by the additional closed channel.

The above analysis motivates a regularization procedure to restore convergence of the iterative Born series.
This is accomplished by spectrally decomposing the kernel operator $\mathbf{K} = \mathbf{K}_R + \mathbf{K}_D$ into a divergent part, $\mathbf{K}_{D}$, which spans the subspace of eigenvectors with $|\eta_{k}|\geq 1$, and a well-behaved regularized part, $\mathbf{K}_{R}$ \cite{Newton_2013}. The divergent part is defined by a separable kernel constructed from the right ($\vec{v}_k$) and left ($\vec{w}_k^{\dagger}$) eigenvectors of $\mathbf{K}$:
\begin{equation}
\label{eq:K_separation}
  \mathbf{K}_D  \vec{u}(r)=    \sum_{k \in \{|\eta_k| \ge 1\}} \eta_k \vec{v}_{k}(r) \int \mathrm{d}r' \vec{w}_{k}^{\dagger}(r') \vec{u}(r') .
\end{equation}
By construction, the spectral radius of the regularized kernel  is less than unity, $\rho(\mathbf{K}_{R}) < 1$. We can therefore reformulate the Lippmann-Schwinger equation to treat the divergent part exactly while solving for the remainder with a convergent Born series. 
To this end, we first define the \textit{regularized} source term, $\vec{u}_{0}^{(R)}$, and the regularized right eigenvectors of $\mathbf{K}$, $\vec{v}_{k}^{(R)}$
\begin{align} \label{eq:regularized_source}
\vec{u}_{0}^{(R)}(r) &= (\mathbf{1} - \mathbf{K}_R)^{-1}\vec{u}_{0}(r), \\
\vec{v}_{k}^{(R)}(r) &= (\mathbf{1} - \mathbf{K}_R)^{-1} \vec{v}_{k}(r), 
\label{eq:regularized_vectors}
\end{align}
where $\mathbf{1}$ is the unit matrix.
These quantities are computed using the  iterative series  based on the regularized  kernel $\mathbf{K}_R$. {\it The regularized series is therefore guaranteed to converge.} The full solution, $\vec{u}(r)$, is then expressed as:
\begin{align}
    \begin{split}
\vec{u}(r) 
 = \vec{u}_{0}^{(R)}(r)
&+ \sum_{k \in \{|\eta_k| \ge 1\}} \eta_k c_{k} \vec{v}_{k}^{(R)}(r),
\label{eq:solution}
    \end{split}
\end{align}
where the coefficients $c_k$ represent the projection of the solution onto the divergent subspace
\begin{equation}
    c_{k} = \int\vec{w}_{k}^{\dagger}({r}')\vec{u}(r')\mathrm{d}r' .
\end{equation}
To determine these coefficients, we project Eq. \eqref{eq:solution} onto the set of left eigenvectors, $\vec{w}_{i}^{\dagger}$.
Crucially, as shown in Fig.~\ref{fig:weinberg}, the number of divergent Weinberg eigenvalues outside the unit circle is much smaller than the total number of eigenvalues of $\mathbf{K}$. The projection thus yields a {\it small} system of linear equations, $\sum_{k \in \{|\eta_k| \ge 1\}} M_{ik}  c_{k} = b_{i}$, where 
\begin{align} 
\begin{split}
\label{eq:linear_eqs} 
b_{i} &= \int\vec{w}_{i}^{\dagger}({r'}) \vec{u}_{0}^{(R)}(r')\mathrm{d}r', \\
M_{ik} & = \delta_{ik} - \eta_{k} \int\vec{w}_{i}^{\dagger}({r'}) \vec{v}_{k}^{(R)}(r') \mathrm{d}r'  .
\end{split}
\end{align}
This framework guarantees convergence regardless of the interaction strength or the presence of closed channels, while retaining the favorable matrix-vector operational scaling. As shown in the right panel in Fig.~\ref{fig:method}(b), the regularization procedure tames the exponential divergence of the Born series for the three-channel model of  He~+~CO collisions with one closed channel, bringing the results in perfect agreement with exact CC calculations.

\subsubsection*{Summary of the algorithm}
\begin{figure}
    \centering
    \includegraphics[width=0.9\linewidth]{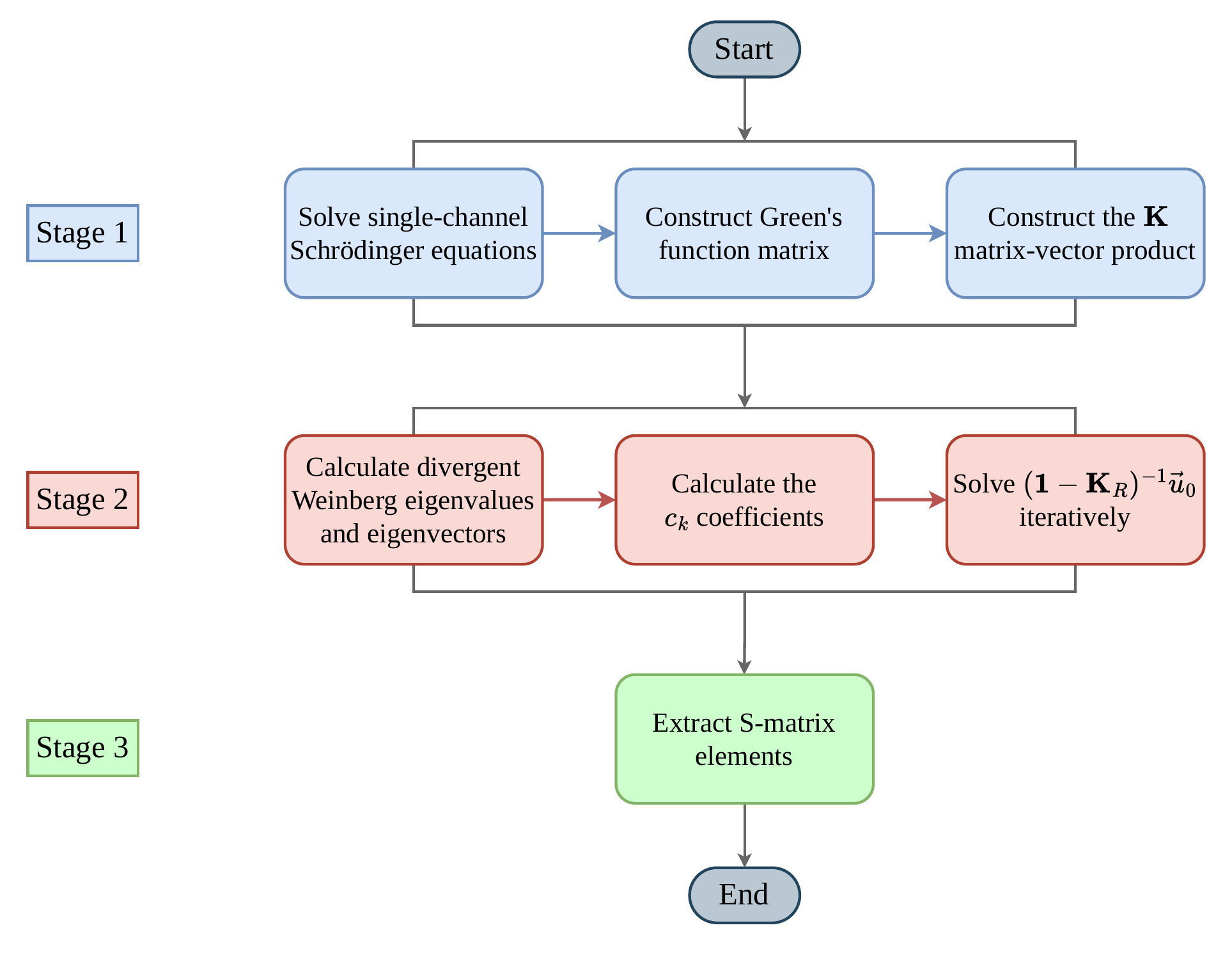}
    \caption{\textbf{Flowchart of the WISE algorithm}. Stage 1 begins by solving a set of independent, single-channel Schrödinger equations to construct the reference Green’s function matrix [Eqs.~\eqref{eq:green_diagonal}-\eqref{eq:green_offdiag}]. This setup enables the efficient computation of matrix-vector products involving the discretized kernel operator $\hat{K}$. Stage 2 constitutes the algorithmic core: the subspace of divergent Weinberg eigenvalues is identified, the regularization procedure is invoked to determine the expansion coefficients $c_k$, and the resulting convergent Born series is summed iteratively. The total scattering wavefunction is then reconstructed via Eq.~\eqref{eq:solution}. Finally, in Stage 3, the scattering matrix elements are extracted from the wavefunction [Eq.~\eqref{eq:Smatrix}].}
    \label{fig:flowchart}
\end{figure}

The practical implementation of the WISE algorithm  proceeds in three distinct stages, as illustrated in Fig.~\ref{fig:flowchart}. First, we construct the reference Green's function by solving a set of independent  single-channel Schrödinger equations  for all channels. Regular and irregular solutions are determined in the form of stable ratios, $Q^{(x)}_{\lambda}(r)$ and $Q^{(y)}_{\lambda}(r)$, by forward and backward propagation, respectively, on the radial grid $r_{i}$. These ratios determine the Green's function matrix via Eqs.~\eqref{eq:green_diagonal} and \eqref{eq:green_offdiag}, enabling the computation of the discretized kernel operator $\mathbf{K}$. Crucially, the algorithm (see Methods) does not require the explicit construction or storage of the full matrices $\mathbf{G}$, $\mathbf{U}^{\mathrm{off}}$, or $\mathbf{K}$. Instead, these are computed on-the-fly, defining the \textit{action} of $\mathbf{K}$ on an arbitrary trial vector as a matrix-vector operation.

Second, we identify the subspace of divergent Weinberg eigenvalues ($|\eta_{k}| \ge 1$) using the Arnoldi iterative algorithm. This requires two passes: first, we compute the $n_{D}$ divergent eigenvalues and their corresponding right eigenvectors, $\vec{v}_k(r)$, by iterating on $\mathbf{K}$. To fully project out the divergent subspace, we also require the left eigenvectors, $\vec{w}_k^{\dagger}$. Because $\mathbf{K}$ is complex and non-symmetric, these are obtained by finding the right eigenvectors of the Hermitian conjugate operator, $\mathbf{K}^{\dagger}$, associated with the conjugate eigenvalues $\eta_k^*$.

We next invoke the regularization procedure by constructing and solving the reduced system of linear equations in Eq.~\eqref{eq:linear_eqs}. This  yields the expansion coefficients  $c_{k}$ necessary to rigorously account for the contribution of the divergent Weinberg eigenvalues, thereby ensuring a convergent iterative series. Because the number of divergent eigenvalues is typically small, this step adds negligible overhead to the overall $\mathcal{O}(MN^2)$ scaling. The terms entering this equation---specifically the action of the regularized source term, $\vec{u}^{(R)}_{0}$, and the regularized right eigenvectors of $\mathbf{K}$, $\vec{v}^{(R)}_k$---are evaluated using the standard Born series, which is now guaranteed to converge. Notably, the evaluation of $\vec{v}^{(R)}_k$ via Eq. \eqref{eq:regularized_vectors} is computationally inexpensive: because $\mathbf{K}_R$ acts as a null operator on the divergent subspace ($\mathbf{K}_R \vec{v}_k \approx 0$), the series truncates immediately. This property implies that the matrix $\mathbf{M}$ in Eq. \eqref{eq:linear_eqs} is predominantly diagonal and highly sparse,  facilitating the use of efficient sparse linear solvers in cases where the dimension of the divergent subspace is large (see Summary and Outlook). With the coefficients $c_{k}$ determined, the full wavefunction is reconstructed via Eq.~\eqref{eq:solution}, and the scattering matrix elements are extracted from Eq.~\eqref{eq:Smatrix} in Stage 3 (see Fig.~\ref{fig:flowchart}), which requires the explicit form of the regular solution $x_{\lambda}(r)$ only for the open channels, consistent with the definition of the  $\mathbf{S}$ matrix.

We note that the divergent eigenvalues and eigenvectors need only to be determined \textit{once per symmetry block} (i.e., for a fixed total angular momentum and spatial parity). Because the kernel $\mathbf{K}$ depends only on the system Hamiltonian and the reference Green's function, its divergent subspace is independent of the incoming channel. Consequently, the iterative procedure can be efficiently applied to all asymptotically open channels within the block. For each new initial state, one simply updates the source term $\vec{u}_{0}$, evaluates its corresponding expansion coefficients $c_{k}$, and computes the action of the regularized kernel $\mathbf{K}_R$ via the Born series.



The Arnoldi-based implementation enjoys $O(N^2)$ scaling in the limit of large $N$, where the number of divergent Weinberg eigenvalues $n_D$ saturates in accordance with the Riesz-Schauder theorem (see Methods for a numerical illustration of the "Riesz-Schauder limit" for He~+~CO). In the pre-asymptotic regime, however, $n_D$ can grow linearly with $N$, and the cost of the Arnoldi step is then $O(n_D × N^2) = O(N^3)$.
To obtain $O(N^2)$ scaling that does not depend on the behavior of $n_D(N)$, we recast the construction of the spectral projector onto the subspace of divergent Weinberg eigenvalues, $\hat{P}_D$ (see Methods), as a matrix-free contour integral \cite{HighamBook,Polizzi:09}. Because the contour $\Gamma_D$ is a fixed boundary enclosing the divergent subspace, evaluated at a fixed number of quadrature nodes with an $O(N^2)$ matrix-vector product, the computational cost is $O(N^2)$ by construction, and is  independent of $n_D$ regardless of whether $n_D$ saturates, grows linearly, 
or is a significant fraction of $N$.
The only quantity that must remain bounded in $N$ is the iteration count of the inner linear solves at each quadrature node; we demonstrate this property directly (see Fig.~9(B) in Methods). Given this bounded-iteration property, the contour-integration construction removes the $n_D$ dependence of the algorithmic complexity and yields $O(N^2)$ scaling for a fixed scattering Hamiltonian.
\color{black}

\subsection*{Application}
The WISE framework is broadly applicable to challenging problems in quantum molecular collision  dynamics, cold and ultracold chemistry, and astrochemistry. To demonstrate its capabilities, we first apply the algorithm to a benchmark system known for its rich resonant structure \cite{Balakrishnan:00,CecchiPestellini:02,Florian:04,Bergeat:15}: cold collisions of CO  molecules with He atoms beyond the reduced-dimensional model described above. Accurate knowledge of low-temperature He~+~CO collision rates is
essential for modeling the chemical properties of the interstellar medium, where CO is a primary tracer molecule \cite{Bergeat:15}.

\begin{figure}
    \centering
    \includegraphics[width=1.00\linewidth]{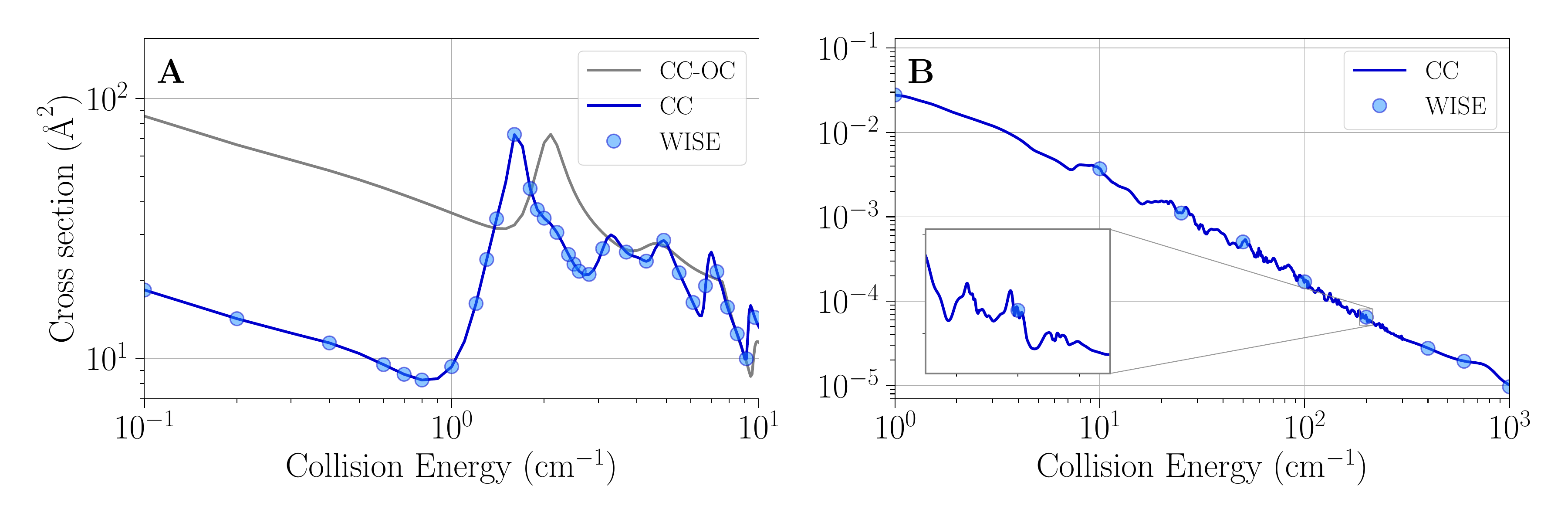}
    \caption{\textbf{Validation and application of the WISE method.} \textbf{(A)} Integral cross-sections for rotational relaxation ($j=1 \to 0$) in He~+~CO collisions plotted as a function of collision energy. The standard CC reference solution (solid blue line) is perfectly reproduced by the WISE method (blue circles). Note that the WISE calculations were performed on a dense energy grid; only a subset of points is shown for clarity. To highlight the physical importance of closed channels, the grey curve shows the results of CC calculations including only open channels, which do not reproduce the  rich Feshbach resonance structure that arises directly from closed-channel couplings. \textbf{(B)} Integral cross-sections for rotational de-excitation of CO ($j_{\mathrm{CO}}=7 \to 6$) in collisions with N$_2$ ($j_{\mathrm{N_{2}}}=6$). This system represents a computationally demanding regime relevant to atmospheric modeling, demonstrating the method's stability for anisotropic molecule-molecule interactions with high channel densities over a broad energy range. }
    \label{fig:COHe}
\end{figure}

Figure~\ref{fig:COHe}(a) shows the total integral cross-section for rotational de-excitation (${j=1 \rightarrow 0}$) in He~+~CO collisions. Our method perfectly recovers the reference CC result, including Feshbach resonances near the channel thresholds \cite{Balakrishnan:00,CecchiPestellini:02,Florian:04,Bergeat:15}. To highlight that we are truly operating in a regime not accessible to previous iterative approaches \cite{Thomas_1979,Thomas_1982}, we additionally present the results obtained with only open channels included. The energy dependence of the ``open-channel-only'' cross-section is smooth and fails to reproduce the resonances, confirming that the inclusion of closed channels is crucial for an accurate description of collision dynamics at the state-to-state, energy-resolved level.

Having validated the WISE approach for rotationally inelastic He~+~CO scattering in the resonant regime, we now demonstrate its broader applicability by addressing a much more complex system: collisions of CO with molecular nitrogen (N$_{2}$). The CO-N$_2$ interaction  is much more anisotropic than that of He-CO 
 \cite{Cybulski_2018},
serving as a stringent test of our algorithm's stability in systems with high densities of states. 
Beyond its theoretical complexity, the CO-N$_{2}$ system is of paramount importance to Earth science. Carbon monoxide is a primary atmospheric pollutant that serves as a crucial tracer of anthropogenic activity, specifically fossil fuel combustion~\cite{Stephens_1991,McMillan_2010,Shan_2019}. Furthermore, while CO is not itself a direct greenhouse gas, it plays a critical indirect role in the climate system: by reacting with hydroxyl radicals, CO depletes the primary sink of methane, thereby increasing the atmospheric lifetime of this potent greenhouse gas~\cite{carole1989global,Daniel_1998}. 
Consequently, global monitoring of CO via satellite remote sensing is essential.

To accurately retrieve CO column densities from spectral observations, one must account for perturbations caused by collisions with dominant atmospheric constituents: N$_{2}$ and O$_{2}$. Specifically, these collisions lead to pressure broadening and shift of the CO spectral lines \cite{Nishiyama_2024}. Modeling these effects from first principles allows for minimizing systematic errors in retrieval algorithms \cite{jacob1999introduction}. The gold standard for such modeling is CC calculations based on high-quality \textit{ab initio} potential energy surfaces \cite{Hartmann2008}. However, the combination of small rotational constants of CO and N$_{2}$, strong potential anisotropy, and the large number of partial waves required at thermal energies, makes fully-converged CC calculations prohibitively expensive. Consequently, first-principles studies of N$_{2}$- and O$_{2}$-perturbed spectra of CO have been limited to only the lowest rotational transitions \cite{Jozwiak_2021,Paredes_Roibas_2021,Zadrozny_2022}.

To explore the robustness of the WISE algorithm when scaled to these computationally demanding regimes,  
we compute inelastic cross-sections for the rotational de-excitation of CO ($j=7 \to 6$) in collisions with N$_{2}$ in its most populated rotational state ($j_{\mathrm{N_{2}}} = 6$). We specifically select this transition because the $j=7$ level corresponds to the peak population of CO at terrestrial temperatures ($\sim300$ K), making it physically representative of the dominant collision dynamics in the atmosphere. 

This setup creates a scattering problem of {substantially}
greater complexity than He~+~CO collisions: the number of channels increases by an order of magnitude. Given the standard cubic scaling, this corresponds to a 1000-fold increase in computational cost.  Figure~\ref{fig:COHe}(b) shows that our WISE algorithm  successfully replicates exact CC results for the de-excitation cross-section across a broad range of collision energies ($1-1000$ cm$^{-1}$). The agreement is excellent over the entire energy range.



\subsection*{Scaling}
 A critical advantage of the WISE method is its favorable scaling with the number of scattering channels, $N$. Upon discretizing the radial coordinate onto a grid, the operator $\mathbf{K}$ becomes a non-symmetric matrix of dimension $(N \cdot N_{\mathrm{g}})\times (N \cdot N_{\mathrm{g}})$. While finding its full spectrum would be computationally prohibitive, our regularization scheme requires only the small subset of divergent eigenvalues with $|\eta_{k}| \geq 1$. These can be found efficiently
 using sparse iterative eigensolvers, such as the Arnoldi algorithm \cite{Arpack}.

 %

To demonstrate the scaling of our algorithm in a realistic setting, we performed a series of benchmarks using the CO-N$_{2}$ system described above. We fixed the collision energy to $10$ cm$^{-1}$ and systematically increased the basis set size, generating a set of coupled equations ranging from $N=12$ to $N=497$.

\begin{figure}
    \centering
    \includegraphics[width=0.99\linewidth]{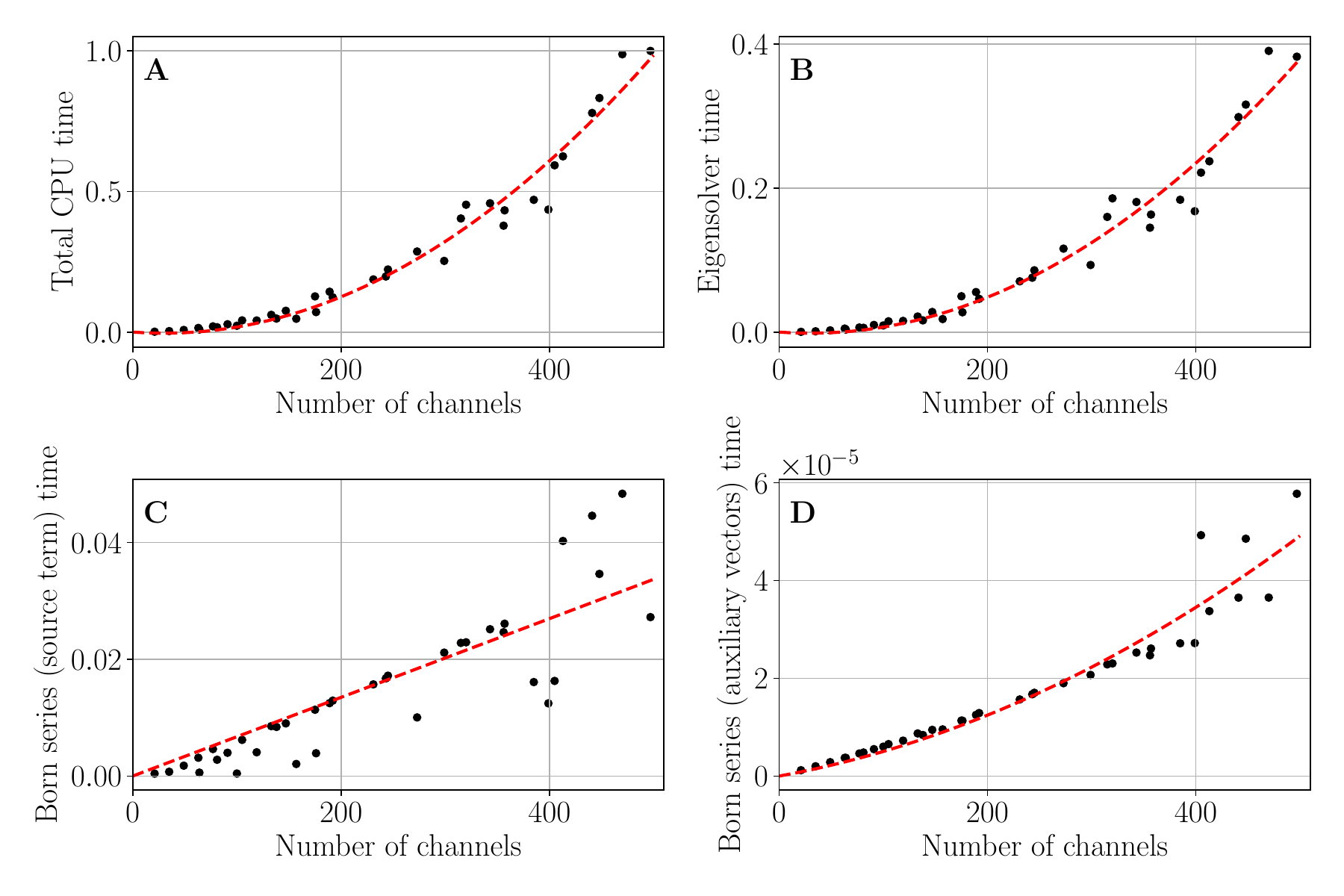}
   \caption{\textbf{Computational scaling of the WISE algorithm.} The wall-clock time (in arbitrary units) is plotted as a function of the total number of scattering channels, $N$. Red dashed lines represent least-squares fits to the data: quadratic ($\propto N^{2}$) for panels (\textbf{A}) and (\textbf{B}), and linear ($\propto N$) for panels (\textbf{C}) and (\textbf{D}). \textbf{(A)} Total time to solution, showing the overall scaling performance. \textbf{(B)} Time required to identify the subspace of divergent Weinberg eigenvalues  with $|\eta_{k}| \ge 1$ using Arnoldi iterations. \textbf{(C)} Time required to converge the Born series for the regularized source vector $\vec{u}^{(R)}_0$. \textbf{(D)} Time required to converge the Born series for the regularized eigenvectors $\vec{v}^{(R)}_k$.}
    \label{fig:scaling}
\end{figure}

The resulting performance is presented in Fig.~\ref{fig:scaling}. The total CPU time shown in Fig.~\ref{fig:scaling}(a) scales quadratically with the number of channels, breaking the cubic bottleneck of the standard approaches based on propagating the scattering wavefunction. This cost is dominated by the iterative search for the divergent eigenvalues and the corresponding right and left eigenvectors [Fig.~\ref{fig:scaling}(b)].

We now consider the computational scaling of the regularization steps.
Figure \ref{fig:scaling}(c) shows the time required to converge the regularized source term, $\vec{u}_{0}^{(R)}(r)$, which typically requires a sequence of multiple iterations.  In contrast, Fig.~\ref{fig:scaling}(d) shows the \textit{cumulative} time to compute the regularized right eigenvectors of  $\mathbf{K}$, $\vec{v}^{(R)}_{k}(r)$, for the entire divergent subspace. Despite involving multiple vectors, this step is extremely fast. Because $\mathbf{K}_{R}$ acts effectively as a null operator on these eigenvectors ($\mathbf{K}_R \vec{v}_k \approx 0$), the Born series truncates almost immediately. Consequently, the scaling observed in this panel is driven not by Born iterations, but rather by the increase in the number of divergent eigenvalues $n_{D}$ with the number of channels $N$. 

Finally, the solution of the sparse linear system for the coefficients $c_{k}$ makes a negligible contribution to the total computational time, even with a standard direct solver used in the present implementation (see Methods). The cost of this step depends only on the number of divergent Weinberg eigenvalues, which can become large ($n_D\simeq 2\times 10^5$) for systems involving an extremely large number of channels ($N\simeq 10^5$), as estimated below. For such systems, the sparse structure of $\mathbf{M}$ [Eq.~(\ref{eq:linear_eqs})] would allow for a further reduction in computational cost using sparse  solvers.

\section*{Discussion}

Quantum CC calculations are an essential tool widely used in atomic, molecular,  chemical, and nuclear physics to elucidate complex scattering phenomena. They are also instrumental in atmospheric chemistry and astrochemistry \cite{Roueff_2013, Lique2019}, underlying quantitative modeling of spectral lineshapes \cite{Hartmann2008,Rayer_2020,Jozwiak2024} and astrochemical reaction networks \cite{Roueff_2013,Dishoeck:13}. All previous algorithms for solving CC equations scale cubically with the number of scattering channels, making these calculations extremely computationally intensive and motivating the development of numerous approximate techniques, ranging from the coupled-states approximation\cite{McGuire:74} to mixed quantum-classical methods \cite{Babikov,Bostan:24}. The accuracy of these approximate techniques is often difficult to estimate. While the need for a practical low-scaling iterative algorithm for solving CC equations  has been recognized for some time \cite{Thomas_1979,Thomas_1982,RhoadesBrown:80,Truhlar:94},  progress has been hindered by the lack of reliable treatment of closed channels and notorious divergence issues with iterative expansions \cite{Thomas_1979,Thomas_1982}.

Here, we overcome these long-standing barriers 
by developing a practical low-scaling algorithm for solving CC equations for a single column of the scattering matrix.  
A key part of the WISE algorithm is the regularization procedure in which the scattering kernel $\mathbf{K}$ is split into the regular and divergent parts using the computed spectrum of Weinberg eigenvalues.  The regular part is treated perturbatively via Born iterations whereas the divergent part is explicitly solved using sparse matrix inversion.
These features ensure numerical convergence of scattering observables and enable the WISE algorithm to readily handle closed channels, making it a promising  tool for currently intractable scattering problems in molecular physics and chemical reactions dynamics.

We have applied the  WISE algorithm to atom-molecule (He~+~CO) and molecule-molecule (CO~+~N$_2$) collisions of relevance to astrochemistry and atmospheric chemistry, demonstrating quadratic scaling (Fig.~\ref{fig:scaling}) and good agreement with benchmark CC calculations (Fig.~\ref{fig:COHe}). While these problems involve  hundreds of coupled channels,
the current limit for conventional algorithms stands at a few tens of thousands of channels
\cite{Suleimanov:12}.

To illustrate the capability of the WISE approach to go beyond this limit, consider, e.g., cold H$_2$O~+~H$_2$O  collisions at interstellar temperatures ($\leq$100~K), which are currently intractable at the exact CC level even in the rigid-rotor approximation. Due to the high anisotropy of the water dimer potential energy surface (PES) \cite{Leforestier:12,Wang:25}, such calculations could easily involve hundreds of thousands of scattering channels.
Using $N= 10^5$ and $100$ optimized radial quadrature points per channel
leads to a 10M $\times$ 10M $\mathbf{K}$-matrix, which can be realistically solved using modern iterative (e.g., Arnoldi) solvers, given the sparse structure of $\mathbf{K}$. The number of Weinberg eigenvalues outside of the unit circle can be estimated as $n_D\leq 10^5$ based on our preliminary CO-N$_2$ calculations. A single WISE calculation would therefore require the inversion of a complex  $n_D\times n_D$  $\mathbf{M}$-matrix. Importantly, this matrix needs to be inverted {\it only once}, as opposed to hundreds of times in conventional CC calculations \cite{Manolopoulos_1986}. Note that the $\mathbf{M}$-matrix is sparse, facilitating the use of highly efficient iterative solvers, such as the generalized minimal residual (GMRES) algorithm  \cite{Manolopoulos:90}.
Using these techniques to extend  the WISE algorithm to $N\simeq 10^5$ is currently in progress.

Finally, we note that  Weinberg eigenvalues can be used as a valuable diagnostic tool for complex multichannel scattering dynamics, as already demonstrated for nucleon-nucleon interactions in  nuclear physics \cite{Hoppe:17}. In particular,  Weinberg eigenvalue spectra  provide  insight into the ``perturbativeness'' of an  interaction PES, i.e., the extent to which scattering observables for the PES are well described by a convergent perturbative expansion. The perturbativeness of a given interaction potential is strongly related to the amount of computational resources required for convergence of scattering observables \cite{Hoppe:17}.
To our knowledge, this work represents the first analysis of Weinberg eigenvalues in the context of multichannel molecular scattering. 
While the physical interpretation of Weinberg eigenvalues merits further study, our calculations show that  they can serve as valuable indicators of strong-coupling phenomena, as their number increases {notably}
in the presence of closed channels and highly anisotropic intermolecular interactions.

While the WISE framework fundamentally improves the algorithmic complexity of the time-independent quantum scattering problem, its current practical implementation presents several specific challenges. When compared to standard CC propagators, which benefited from decades of low-level  optimization, the integral equation approach introduces a different demand on computational resources. Because the WISE approach evaluates the Lippman-Schwinger equation over the entire spatial domain rather than propagating the differential solution sector-by-sector, it intrinsically requires large memory storage. Constructing the Green's function on the fly necessitates storing the ratios of the regular and irregular solutions at every point on the spatial grid. Furthermore, the Arnoldi spectral regularization procedure requires keeping $n_{D}$ left and right eigenvectors of $\mathbf{K}$ (each of size $N\times N_{g}$) in active memory.


We found that using a sufficiently high density of radial grid points in the short-range region provides multiple advantages by (i)  ensuring an accurate representation of the interaction PES; (ii)  
accurately resolving the oscillatory behavior of the short-range Green's function, and (iii)  reducing the number of diverging Weinberg eigenvalues, $n_{D}$. To fully exploit these advantages  without inflating the overall memory cost, future implementations will benefit from the use of optimized spatial grids, such as Gauss-Lobatto quadrature \cite{Manolopoulos:90} and finite-element discrete-variable representations \cite{Rescigno:00} to minimize the total number of grid points $N_{g}$.

Finally, isolating the divergent subspace introduces its own challenges. While the fundamental guarantee of the $O(N^2)$ asymptotic scaling relies on the Riesz-Schauder theorem for compact operators \cite{ReedSimonBook}, which dictates that the number of divergent Weinberg eigenvalues $n_{D}$ must eventually saturate, 
{
the dependence of $n_D$ on the number of channels $N$ and total angular momentum $J$ is generally  non-trivial (see Methods). While there exists an  intermediate stage where $n_D$ scales linearly with $N$, we identify other important regimes where $n_D$ saturates, and even decreases well before achieving convergence. In the specific regime where $n_D=a\times N$, standard Arnoldi iterative solvers exhibit unfavorable cubic scaling, which can be overcome using  matrix-free spectral projection operators constructed via contour interaction (see Methods).}




\section*{Methods}
\subsection*{Potential Energy Surfaces and Hamiltonians}
For the CO-He system, we utilize the \textit{ab initio} PES developed by Peterson and McBane~\cite{Peterson_2005} at the CCSD(T) level of theory. We employ the effective 2D potential calculated for the ground vibrational state ($v=0$) of CO, denoted as $V_{v=0}(r, \theta)$, where $r$ is the distance between the CO center of mass and the He atom, and $\theta$ is the Jacobi angle. The angular dependence of the CO-He PES is expanded in a basis of 20 Legendre polynomials. The internal Hamiltonian $\hat{H}_{\mathrm{int}}$ describes the CO molecule as a rigid rotor:
\begin{equation}
\label{eq:rigid_rotor}
\hat{H}_{\mathrm{CO}} = B_{e} \hat{j}^{2} - D_{e} \hat{j}^{4},
\end{equation}
with the rotational constant $B_e = 1.92251$ cm$^{-1}$ and the  centrifugal distortion constant $D_e = 6.1193 \times 10^{-6}$ cm$^{-1}$~\cite{Huber1979}.

For the CO-N$_2$ system, we employ the 4D PES computed by Cybulski \textit{et al.}~\cite{Cybulski_2018} using the CCSD(T) method with an aug-cc-pVQZ basis set augmented with midbond functions. The intramolecular bond lengths were frozen at their vibrationally-averaged ground-state values ($r_{\mathrm{CO}} = 2.13201,a_{0}$ and $r_{\mathrm{N_{2}}} = 2.07397\,a_{0}$). The full interaction potential is expanded in bispherical harmonics as detailed in Ref.~\cite{Jozwiak_2021}. The internal Hamiltonian is the sum of two rigid rotor Hamiltonians, $\hat{H}_{\mathrm{int}} = \hat{H}_{\mathrm{CO}} + \hat{H}_{\mathrm{N}_2}$, with the rotational constants of CO and N$_2$ taken from Ref. \cite{Huber1979}.

\subsection*{Basis Sets and Channel Definitions}
For the CO-He system, calculations are performed in the space-fixed basis set defined by the coupling of the rotational angular momentum $\hat{j}$ with the end-over-end orbital angular momentum $\hat{L}$ to form the total angular momentum $\hat{J}$. The internal index used in Eq.~\eqref{eq:Psi} is $\gamma \equiv \{j\}$. 

For the reduced-dimensional model (Fig.~\ref{fig:COHe}), calculations were restricted to the $J=0$ block at a collision energy of $E_{\mathrm{kin}} = 5$ cm$^{-1}$. The two-channel model includes only the open channels ($j=0, L=0$ and $j=1, L=1$). The three-channel model adds the lowest closed channel ($j=2, L=2$).
The fully converged CO-He calculations presented in Fig.~\ref{fig:COHe} used an extended basis set including all rotational states up to $j^{\mathrm{max}} = 6$ and total angular momenta up to $J^{\mathrm{max}} = 9$, in addition to all the allowed $L$ values.

To explicitly verify that the WISE algorithm converges with respect to the basis size exactly as standard CC calculations do, we track values of the scattering matrix as closed channels are incrementally added. The top panel of Fig.~\ref{fig:conv} shows the results
for the $j=1 \to 0$ de-excitation transition in He~+~CO collisions at a collision energy of 5 cm$^{-1}$. As a representative example, we examine the $J=0$ block of positive parity $p=(-1)^{j+L}=+1$. Under these conditions, only two channels are asymptotically open: $j=0, L=0$ (denoted as channel $0$), and $j=1, L=1$ (denoted as channel $1$). We plot the squared moduli of the elastic ($|S_{11}|^{2}$) and inelastic ($|S_{01}|^{2}$) matrix elements as a function of the total number of channels $N$ kept in the basis. The sequence begins at $N=2$ (open channels only) and adds closed channels sequentially up to $j=19, L=19$. We observe that the WISE results perfectly trace the benchmark CC values throughout the entire convergence pattern. The iterative scheme exhibits no instabilities or loss of accuracy even in the presence of a large number of closed channels: at $N=20$, where the ratio of closed to open channels is $9:1$, the WISE and CC results are exactly the same.

In the case of the CO-N$_2$ system, the basis is formed by coupling the rotational angular momenta of CO ($j_{\mathrm{CO}}$) and N$_2$ ($j_{\mathrm{N}_2}$) to a resultant $j_{{AB}}$, which is then coupled with $L$ to form $J$. The internal index $\gamma \equiv \{j_{\mathrm{CO}}, j_{\mathrm{N}_2}, j_{AB}\}$. The cross-sections for the $j_{\mathrm{CO}}=7 \to 6$ transition in collisions with N$_2(j_{\mathrm{N}_2}=6)$ were computed for $J=0$ with the basis set that included $j_{\mathrm{CO}} = 0-14$ and $j_{\mathrm{N}_2} = {4, 6, 8}$ (note that coupling between even and odd rotational levels of a homonuclear molecule via the CO-N$_{2}$ PES is forbidden due to nuclear spin symmetry). This results in a system of 248 (even parity) and 203 (odd parity) CC equations, ensuring convergence of the de-excitation cross-section to within 20\%.

To analyze the computational scaling, we performed a series of CO-N$_{2}$ calculations at a fixed collision energy of 10 cm$^{-1}$. We defined a minimal basis set containing only the initial and final states ($j_{\mathrm{CO}} \in {6,7}, j_{\mathrm{N}_{2}}=6$) and systematically expanded the number of channels by adding the rotational states of both monomers (expanding to $j{_\mathrm{CO}} < 6, j_{\mathrm{CO}} > 7$ and $j_{\mathrm{N}_2} \neq 6$). This procedure generates a set of scattering problems with total channel counts ranging from 12 to 497.

\subsection*{Numerical Implementation of the Iterative Algorithm}
The radial grid and basis sets used in the WISE calculations are identical to those employed in the benchmark CC production runs. For He~+~CO collisions, the kernel matrix (Eq.~\ref{eq:K_matrix}) is discretized on a radial grid spanning $r \in [3.4, 20.0]\,a_{0}$ with $N_{\mathrm{g}} = 1661$ points ($\Delta r = 0.01\,a_{0}$). For CO~+~N$_2$ collisions, we use a grid of $N_{\mathrm{g}} = 1581$ points ranging from $4.2$ to $20.0\,a_{0}$ ($\Delta r = 0.01\,a_{0}$), which ensures strict convergence ($<1\%$) of benchmark CC calculations  \cite{Johnson_1978}.
{For cold and ultracold collisions, 
the outer range
of radial integration may have to be extended to $\simeq$100 $a_0$ or longer, which can be readily handled by  specialized Gauss-Lobatto or finite-element methods.}
The quadrature weights $w_{i'}$ entering the definition of the discretized $\mathbf{K}$ matrix in Eq.~\eqref{eq:K_matrix} follow the composite trapezoidal rule: $w_{i}=\Delta r$ for internal points $i\in \{2, N_{\mathrm{g}}-1\}$ and $w_{i}= {\Delta r} / {2}$ for the endpoints $i\in \{1, N_{\mathrm{g}}\}$. 

A critical feature of the WISE algorithm is that the full kernel matrix $\mathbf{K}$ is never explicitly constructed or stored in memory, which would otherwise impose a prohibitive $\mathcal{O}(N^2 N_{\mathrm{g}}^2)$ memory cost. Instead, we treat $\mathbf{K}$ as a matrix-free linear operator, defined solely by its action 
on a trial vector. In our implementation, this is achieved using the \texttt{LinearOperator} abstraction in SciPy, which interfaces directly with sparse eigensolvers. The subset of eigenvalues required for regularization is computed using the Implicitly Restarted Arnoldi Method, as implemented in the ARPACK library \cite{Arpack}. This allows us to selectively converge only the divergent eigenvalues with magnitudes $|\eta| \ge 1$, avoiding the cost of a full spectral decomposition.

Once the divergent subspace is identified, the regularization coefficients $c_{k}$ are determined by solving a system of linear equations \eqref{eq:linear_eqs} of dimension $n_{D} \times n_{D}$, where $n_{D}$ is the number of divergent Weinberg eigenvalues. Since the system is sparse, and the set of equations has to be solved only once, the total computational cost introduced by this stage is negligible. The set of equations is solved using the standard LU decomposition driver for general complex matrices ($\texttt{ZGESV}$) from the LAPACK library, accessed via NumPy's linear algebra wrapper.

The Born series for the regularized remainder is evaluated via fixed-point iteration. At each step $n$, the solution is updated as $\vec{u}^{(n)} = \vec{u}_{0} + \mathbf{K}_{R}\vec{u}^{(n-1)}$, and the single column of the $S$-matrix is computed according to Eq. \eqref{eq:Smatrix}. Convergence is monitored by explicitly tracking the maximum absolute change in the scattering matrix elements between consecutive iterations, $\epsilon = \mathrm{max}|S_{ij}^{(n)} - S_{ij}^{(n-1)}|$. The iterations are terminated when $\epsilon$ falls below a specified tolerance threshold, set to $10^{-6}$ in the present work. 

Because the spectral radius of the regularized operator $\rho(\mathbf{K}_{R})$ is strictly less than 1, this iterative sequence is formally guaranteed to~converge. However, the practical speed of convergence is governed by the largest remaining Weinberg eigenvalues. Eigenvalues with magnitudes strictly less than, but close to, unity can 
slow down the convergence of the Born series. This bottleneck can be easily circumvented by extending the regularized subspace to project out not only the strictly divergent eigenvalues ($|\eta|>1$), but also those that merely impede convergence. In the present calculations leading to Figures \ref{fig:COHe} and \ref{fig:scaling}, we optimize this by setting the exclusion threshold to $|\eta| \geq 0.9$.

The efficiency of this optimized regularization scheme is illustrated in Fig. \ref{fig:M_vs_N}, which presents the number of iterations $M$ required to converge the single column of the $S$-matrix as a function of the number of channels, $N$ for the N$_{2}$+CO calculations used to demonstrate scaling of the WISE algorithm on Fig. \ref{fig:scaling}. While there is no strictly monotonic pattern ($M$ is scattered between 125 and 225, depending on the specific spectrum of $\mathbf{K}$ for a given set of channels), it is clear that $M$ does not grow continuously with $N$. For a small number of channels, the number of iterations can exceed the basis size ($M> N$), making the iterative approach less practical for simple systems. Crucially, however, as the basis set size increases, the required number of iterations saturates. For the largest number of channels considered here ($N=497$), convergence is achieved with $M\approx 218$ steps, clearly demonstrating that in the asymptotic limit of large $N$, the iterative part of the algorithm operates within the $M<N$ regime.

\subsection*{Weinberg eigenvalues: Scaling with the number of channels}

The number of iterations required by the Arnoldi eigensolver scales linearly with the number of targeted eigenvalues. Consequently, if the number of diverging eigenvalues, $n_{D}$, were to grow linearly with the number of channels, $N$, the computational cost of the Arnoldi step would scale as $\mathcal{O}(n_{D}\times N^{2}) = \mathcal{O}(N^{3})$. While the Riesz-Schauder theorem formally guarantees that $n_{D}$ must eventually saturate at large $N$, there can exist an intermediate regime where $n_{D}$ grows linearly, temporarily manifesting unfavorable cubic scaling. However, as $N$ increases further, the growth of $n_{D}$ is mathematically forced to become sublinear and ultimately approach a constant.

\begin{figure}
    \centering
    \includegraphics[width=0.5\linewidth]{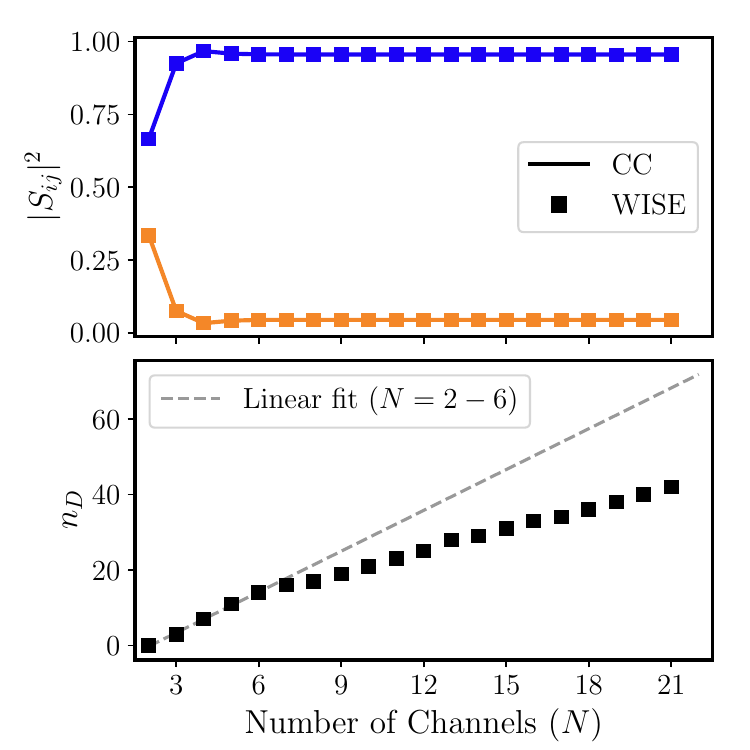}
\caption{\textbf{Convergence of the WISE algorithm with respect to the number of scattering channels.} (Top) Convergence of the transition probabilities ($|S_{ij}|^2$) as a function of the number of channels, $N$, for He~+~CO scattering at a collision energy of $5$ cm$^{-1}$ ($J=0, p=1$). The plot displays both the elastic ($|S_{11}|^2$) and inelastic ($|S_{01}|^2$) transitions between the two asymptotically open channels: $j=0, L=0$ (channel $0$) and $j=1, L=1$ (channel $1$). Solid lines indicate the benchmark values obtained using standard coupled-channel (CC) calculations, while the symbols represent the results computed using the WISE algorithm. (Bottom) Number of diverging Weinberg eigenvalues, $n_{D}$, as a function of the total number of channels for the same system. The dashed line highlights the regime in which $n_{D}$ scales linearly with $N$.}    
    \label{fig:conv}
\end{figure}

The bottom panel of Fig.~\ref{fig:conv} illustrates this behavior, showing the dependence of $n_{D}(N)$ for
He~+~CO collisions at a collision energy of $5$ cm$^{-1}$  for the $J=0$, $p=+1$ symmetry block. We observe a linear growth of $n_{D}$ up to $N=6$, after which the trend becomes sublinear. Interestingly,{in this specific example,} the number of channels at which this scaling transitions coincides with the physical convergence of the $S$-matrix elements, as seen in the top panel of Fig.~\ref{fig:conv}.

\begin{figure}
    \centering
    \includegraphics[width=0.92\linewidth]{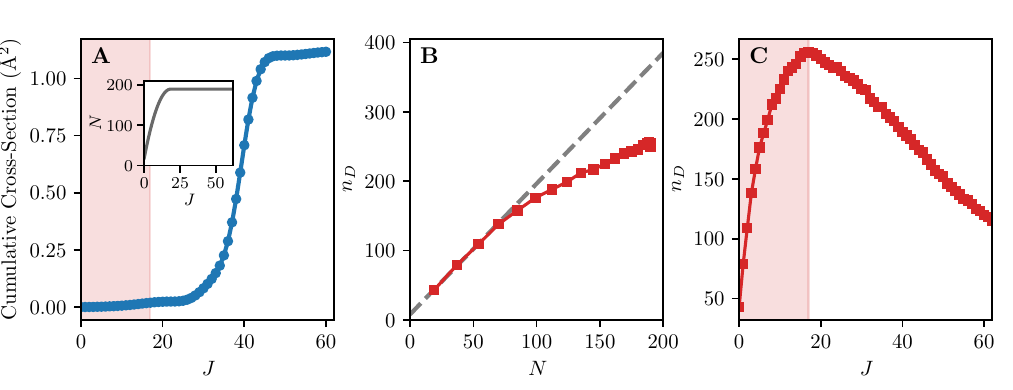}
\caption{\textbf{Convergence of the WISE algorithm with respect to the total angular momentum  $J$ of the collision complex.} {\bf (A)} Cumulative cross-section as a function of $J$, which reaches convergence at $J \approx 50$. {\bf (B)} The number of diverging Weinberg eigenvalues as a function of the number of coupled channels ($N$). The growth becomes sub-linear around $N=100$ ($J = 5$). {\bf (C)} The number of diverging Weinberg eigenvalues as a function of $J$. The number peaks at $n_D=256$ ($J = 17$, $N=189$). Increasing $J$ beyond this point does not increase the number of channels (which saturates at $N=190$), but rather actively {decreases} the number of diverging Weinberg eigenvalues. The red shaded area in panels {\bf (A)} and {\bf (C)} corresponds to the range of $J=0-18$ values presented in panel~{\bf (B)}.}    
    \label{fig:totalJconvergence}
\end{figure}

\begin{figure}
    \centering
    \includegraphics[width=0.5\linewidth]{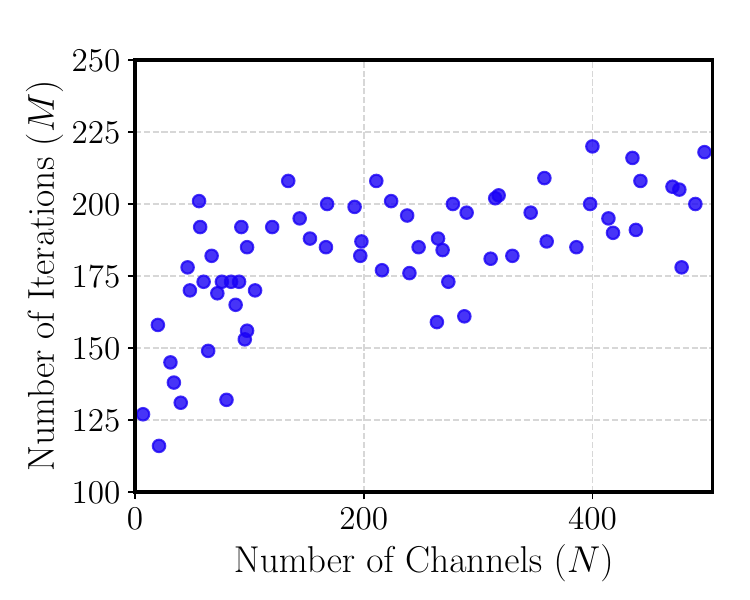}
    \caption{\textbf{Saturation of the iteration count in the large basis limit.} Number of iterations $M$ required to converge the regularized Born series as a function of the number of coupled channels $N$. The dataset corresponds to the N$_{2}$+CO scaling tests presented in Fig. \ref{fig:scaling}. By extending the regularized subspace to include eigenvalues with magnitudes $|\eta| \geq 0.9$, the physical $S$-matrix converges to a tolerance of $10^{-6}$ in approximately $120-225$ iterations, demonstrating that $M$ saturates and becomes decoupled from $N$ in the large basis limit.}
    \label{fig:M_vs_N}
\end{figure}

However, the dependence $n_D(N)$ can also exhibit a range of other behaviors, including sub-linear growth. One realistic scenario is studying the convergence of scattering observables as a function of the total angular momentum $J$. Indeed, converging $j_{\mathrm{max}}$ within a single $J$ block is only part of solving the actual scattering problem; a converged
cross section requires solving the CC equations across a range of $J$ values.
To examine how the problem size and computational complexity evolve in such a scenario,  we plot in Fig.~\ref{fig:totalJconvergence} the convergence of the cross-section for rotational relaxation ($j=1\to 0$) in He~+~CO collisions at a collision energy of $800\,\mathrm{cm}^{-1}$. 
Panel~(A) shows the cumulative cross-section as a function of  $J$, reaching convergence around $J \approx 50$. The inset of Panel (A) details how the physical size of the problem (the number of channels, $N$) changes with $J$. Initially, as $J$ increases, the space-fixed basis expands to accommodate higher orbital angular momentum states ($L$). However, once $J$ exceeds the maximum rotational state in the basis ($j_{\mathrm{max}} = 18$ in this case) the number of channels saturates at $N=190$ due to triangular angular momentum selection rules.



Based on the $J = 0$ tests shown in the lower panel of Fig.~\ref{fig:conv}, one might expect $n_{D}$ to follow the same pattern: growing linearly with $N$ and then remaining constantly high once $N$ saturates.
Contrary to this expectation,  $n_D$ does not remain high: as a function of $J$
it rises to a maximum of 256 at $J = 17$ and then decreases [Fig.~\ref{fig:totalJconvergence}(C)], and as a function of
$N$ its growth becomes markedly sub-linear above $N \approx 100$, long before the cross section reaches convergence [Fig.~\ref{fig:totalJconvergence}(B)]. 
While this empirical, system-specific observation  indicates that the aggregate cost of a converged cross section can be well below the worst-case fixed-$J$ estimate, we
do not present it as a proof of algorithmic scaling. We also note  that at its 
peak value, $n_D = 256$ is of order $N$.


\color{black}

\subsection*{Matrix-free spectral projection via contour integration}
To avoid the cubic scaling of the diagonalization step in the transient pre-asymptotic regime, where $n_D=a\times N$ (see above), we reformulate the WISE framework to bypass the explicit evaluation of the individual Weinberg eigenvalues. Instead, we perform spectral projection via a contour integral approach.  Using the operator identity
$(\hat{A} - \hat{B})^{-1} = A^{-1} + (\hat{A} - \hat{B})^{-1} \hat{B} \hat{A}^{-1}$ which can be  verified by multiplying both sides by $(\hat{A}-\hat{B})$ on the left, and setting $\hat{A}=\hat{I}-\hat{K}_R$ and $\hat{B}=\hat{K}-\hat{K}_R=\hat{K}_D$, where $\hat{I}$ is the identity operator, we obtain
\begin{equation}\label{eq:operator_id}
(\hat{I} - \hat{K})^{-1} = (\hat{I} - \hat{K}_{R})^{-1} + (\hat{I} - \hat{K})^{-1}  \hat{K}_D(\hat{I} - \hat{K}_{R})^{-1}.    
\end{equation}
Here, $\hat{K}=\hat{K}_R + \hat{K}_D$ is the LS kernel operator, which corresponds to the matrix $\mathbf{K}$ defined in the main text, partitioned into the regularized and divergent parts.
Applying Eq.~\eqref{eq:operator_id} to the incident state $|u_0\rangle$, we obtain the exact solution of the LS equation as
\begin{equation}\label{eq:LS_mod}
|u\rangle = (\hat{I} - \hat{K})^{-1}|u_0\rangle = (\hat{I} - \hat{K}_{R})^{-1}|u_0\rangle + (\hat{I} - \hat{K})^{-1}  \hat{K}_D(\hat{I} - \hat{K}_{R})^{-1}|u_0\rangle.    
\end{equation}
Recognizing $(\hat{I} - \hat{K}_{R})^{-1}|u_0\rangle=|u_0^{(R)}\rangle$ as the regularized source term in Eq.~\eqref{eq:regularized_source}, we can recast Eq.~\eqref{eq:LS_mod} as
\begin{equation}\label{eq:LS_mod2}
|u\rangle = |u_0^{(R)}\rangle + (\hat{I} - \hat{K})^{-1}  \hat{K}_D |u_0^{(R)}\rangle.    
\end{equation}
This result is identical to Eq.~\eqref{eq:solution}, with  the second term now expressed in operator form. 

Our goal is to evaluate the regularized and divergent contributions  in Eq.~\eqref{eq:LS_mod2} without explicitly computing the spectrum of divergent Weinberg eigenvalues. 
To this end, we define the spectral projector $\hat{\mathcal{P}}_D$ onto the divergent subspace as a contour integral in the complex plane \cite{HighamBook, Hale:08}
\begin{equation}\label{eq:countour_int}
\hat{\mathcal{P}}_D  = \frac{1}{2\pi i} \oint_{\Gamma_D}  (z\hat{I} - \hat{K})^{-1} dz,
\end{equation}
where $\Gamma_D$ is a closed ring contour enclosing the  manifold of divergent Weinberg eigenvalues. The contour contains a narrow cut connecting the inner and outer circles, which could ``miss'' divergent eigenvalues if their density becomes very high. In this case, a scaling transformation  $\hat{K}\to \hat{K}^{-1}$ could be applied to map the diverging Weinberg eigenvalues onto the interior of the unit circle which becomes the new integration contour.
 With the definition \eqref{eq:countour_int}, we have  $\hat{K}_D=\hat{K}\hat{\mathcal{P}}_D$ and   $\hat{K}_R=\hat{K}(\hat{I}-\hat{\mathcal{P}}_D)$. Using the expression  $f(\hat{A})=\frac{1}{2\pi i}\oint_{\Gamma_A} f(z)(z\hat{I}-\hat{A})^{-1}dz$ \cite{Hale:08}, where $\hat{A}$ is an operator,   $f(z)$ is an analytic function, and the contour $\Gamma_A$ encloses the spectrum of $\hat{A}$,  Eq.~\eqref{eq:LS_mod2} may be written as
 \begin{equation}\label{eq:LS_mod3}
 |u\rangle = |u_0^{(R)}\rangle + \frac{1}{2\pi i} \oint_{\Gamma_D} \frac{z}{1-z} (z\hat{I} - \hat{K})^{-1} |u_0^{(R)}\rangle dz,
 \end{equation}
 where we have used the fact that $\hat{\mathcal{P}}_D$ commutes with $\hat{K}$.
  The first term can be  expressed as a convergent Born series   $|u_0^{(R)}\rangle=(\hat{I}-\hat{K}_R)^{-1} |u_0\rangle=\sum_{n}\hat{K}_R^n |u_0\rangle$ with $\hat{K}_R|u_0\rangle=\hat{K} |u_0\rangle - \hat{K} \hat{\mathcal{P}}_D|u_0\rangle = \hat{K} |u_0\rangle - \frac{1}{2\pi i} \oint_{\Gamma_{D}}  z (z\hat{I} - \hat{K})^{-1} |u_0\rangle \, dz$.



 Importantly, the contour integral representation \eqref{eq:LS_mod3}  {\it guarantees $O(N^2)$ scaling regardless of the dependence $n_D(N)$}
because all diverging Weinberg eigenvalues can be enclosed within a fixed-size contour $\Gamma_D$ determined by the largest eigenvalue $\eta_\text{max}$, which can be computed in  $O(N)^2$ operations.
Therefore, it is no longer necessary to compute the individual eigenvalues to evaluate the action of the projector $\hat{\mathcal{P}}_D$ on an arbitrary vector $|v\rangle$. Instead, we evaluate the Cauchy integral in Eq.~\eqref{eq:LS_mod3} via numerical quadrature [with $|v\rangle=|u_0^{(R)}\rangle$ for the second term in Eq.~\eqref{eq:LS_mod3}]
\begin{equation}\label{countour_int_quadrature}
\oint_{\Gamma_D} \frac{z}{1-z} (z\hat{I} - \hat{K})^{-1} |v\rangle dz \approx \sum_{j=1}^{N_q} w_j \frac{z_j}{1-z_j} [(z_j\mathbf{1} - \mathbf{K})^{-1}\vec{v}] = \sum_{j=1}^{N_q} w_j  \frac{z_j}{1-z_j} \vec{y}_j 
\end{equation}
 by solving the linear system $(z_j \mathbf{1} - \mathbf{K}) \vec{y}_j = \vec{v}$
for a small, fixed number of quadrature points $z_j$ (typically $N_q \simeq 16$–$32$)   
 using a quadratically scaling iterative solver such as GMRES \cite{Manolopoulos:90}.
Because our contour $\Gamma_D$ is a fixed physical boundary independent of $n_D$,  the accuracy of the quadrature approximation \eqref{countour_int_quadrature} depends on the analyticity of the resolvent along the contour rather than the density of the poles it encloses \cite{Trefethen:14,Polizzi:09}.
This allows the WISE framework to  decouple the computational cost from the subspace rank $n_D$, maintaining $O(N^2)$ scaling even in regimes where $n_D$ scales linearly with $N$.

{We have implemented the matrix-free spectral projection approach for He~+~CO collisions using the same framework as described above  (see ``Numerical Implementation of the Iterative Algorithm'' in Methods). The integral kernel is discretized on the uniform radial grid using trapezoidal quadrature weights. The kernel matrix $\mathbf{K}$ is implemented as a matrix-free linear operator defined solely by its action on an arbitrary vector, using the \texttt{LinearOperator} abstraction in SciPy.}

\begin{figure}
    \centering
    \includegraphics[width=\linewidth]{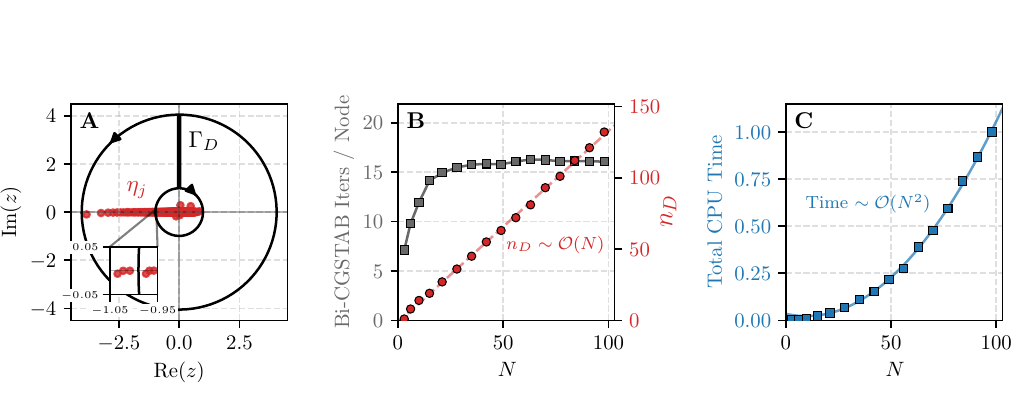}
    \caption{{\textbf{Matrix-free spectral projection and computational scaling.} \textbf{(A)} Example integration contour ($\Gamma_D$) in the complex plane enclosing the divergent Weinberg eigenvalues ($\eta_j$) for the He + CO collision system for the total angular momentum and parity $J=6$, $p=1$ at a collision energy of $E=5\,\mathrm{cm}^{-1}$. The inset shows  Weinberg eigenvalues in the vicinity of the inner ring of the contour. \textbf{(B)} Average number of Bi-CGSTAB iterations per quadrature node (left axis, gray squares) required to solve the linear system of equations [Eq. \eqref{countour_int_quadrature}] as a function of the basis size ($N$). The secondary axis (right axis, red circles) tracks the number of divergent Weinberg eigenvalues ($n_D$), which scales linearly with $N$. \textbf{(C)} Total wall-clock time (in arbitrary units, blue squares) for the same scattering system.}}
    \label{fig:contour_integral}
\end{figure}

{The contour integrals  in Eq.~\eqref{eq:LS_mod3} are evaluated  by numerical quadrature. The integration path $\Gamma_D$ is chosen as an annular contour consisting of a large outer ring traced counter-clockwise, connected by a branch cut along the positive imaginary axis, as shown in Figure \ref{fig:contour_integral}(A).
The outer ring is defined by the Weinberg eigenvalue of largest magnitude, $|\eta_{\max}|$, computed using the Implicitly Restarted Arnoldi Method (as implemented in ARPACK). Because we only require a single extremal eigenvalue, {this step needs a small number of Arnoldi iterations in practice, each costing $O(N^2)$.} The outer radius is then set to $R_{\text{out}} = |\eta_{\max}| + \delta$, where $\delta = 0.2$ provides a safe numerical buffer, and the integral over this outer boundary is discretized with trapezoidal quadrature points ($N_{q}^{\text{out}} \approx 30$). The inner ring is centered at the origin with a radius $R_{\text{in}} \approx 1$ to enclose the divergent eigenvalues ($|\eta_{j}| \geq 1$). At each quadrature point $z_k$, the shifted linear system $(z_k\mathbf{I} - \mathbf{K})\vec{y}_k = \vec{v}$ is solved using the Biconjugate Gradient Stabilized (Bi-CGSTAB) method \cite{van_der_Vorst_1992}.}

{Figures~\ref{fig:contour_integral}(B) and (C) show the results of the contour integration approach for a representative test case of rotational relaxation ($j=1\to 0$) in He~+~CO collisions at a collision energy of $E=5\,\mathrm{cm}^{-1}$. Specifically, we solve the CC equations for the total angular momentum and parity block $J=6$, $p=1$, and the incoming channel $j=1, L=5$. The basis set size $N$ is systematically expanded by increasing the maximum rotational state $j_{\max}$. As shown in panel (B), while the number of divergent Weinberg eigenvalues ($n_D$, red circles, secondary axis) grows linearly with $N$ in this pre-saturation regime, the corresponding number of Bi-CGSTAB iterations (gray squares) saturates. Consequently, the total CPU time maintains quadratic $\mathcal{O}(N^2)$ scaling, as shown in Fig.~\ref{fig:contour_integral}(C).}

As is standard in contour-based spectral projection techniques, the convergence rate of the iterative linear solver depends on the proximity of the quadrature nodes $z_k$ to the eigenvalues \cite{Kestyn:16}.
While an unoptimized inner contour may occasionally place a quadrature node near a diverging Weinberg eigenvalue, the Riesz-Schauder theorem guarantees that eigenvalues of a compact operator accumulate exclusively at the origin ($z=0$). Consequently, the spectrum at the inner contour boundary ($|z| \approx 1$) is strictly finite and discrete. This discreteness guarantees the existence of empty spectral gaps [inset,    Fig.~\ref{fig:contour_integral}(A)], 
allowing small perturbations of the inner radius or quadrature-point phases to bypass localized poles even as $n_D$ grows. For the present He~+~CO system, the divergent eigenvalues lie close to the negative real axis [inset, Fig. 9(A)], so the inner contour crosses them in a confined, well-gapped region; the inner solves correspondingly maintain a bounded iteration count and a well-conditioned shifted system [Fig.~9(B)]. The behavior at the much higher densities relevant to H$_2$O + H$_2$O collisions is the subject of work in progress.


\color{black}

\newpage

\color{black}

\color{black}



\clearpage 

%
\bibliography{science_template} 
\bibliographystyle{sciencemag}


\noindent{\bf Acknowledgments:}
We thank George McBane and Jeremy Hutson for stimulating discussions in the early stages of this work and Nathan Prins for valuable comments.

\noindent{\bf Funding:}
This work was supported by the National Science Centre in Poland through Project No. 2024/53/N/ST2/02090 (HJJ) and 
by the NSF CAREER award No. PHY-2045681 (MMR and TVT).

\noindent{\bf Author Contributions:}
Conceptualization: HJJ, TVT;
Methodology: HJJ, TVT;
Investigation: HJJ, MMR; 
Visualization: HJJ, MMR;
Supervision: TVT, HJJ;
Writing—original draft: HJJ, TVT;
Writing—review \& editing: HJJ, TVT, MMR.

\noindent{\bf Competing Interest:} The authors declare they have no competing interest.

\noindent{\bf Data, Code, and Materials Availability:} 
All data and code needed to evaluate and reproduce the results in the paper are present in the paper and/or the Supplementary Materials and are available online: \url{https://doi.org/10.5281/zenodo.19395019}. This study did not generate new materials.

\end{document}